\documentclass[%
 reprint,
 amsmath,amssymb,
 aps,
]{revtex4-1}

\usepackage{graphicx, caption}
\usepackage{float}
\usepackage{caption,subcaption}
\usepackage{lipsum}
\usepackage{dcolumn}
\usepackage{bm}
\usepackage{xcolor} 

\begin{document}
\newcommand{\bra}[1]{\langle#1\rvert} 
\newcommand{\ket}[1]{\lvert#1\rangle} 

\title{Confined magnon dispersion in ferromagnetic and antiferromagnetic thin films in a second quantization approach: the case of Fe and NiO}

\author{Julio A. do Nascimento}
 \altaffiliation{julio.nascimento@york.ac.uk}
 \affiliation{School of Physics, Engineering and Technology, University of York, York YO10 5DD, UK}
 
\author{S. A. Cavill}%
\affiliation{School of Physics, Engineering and Technology, University of York, York YO10 5DD, UK}%
\author{Adam Kerrigan}%
\affiliation{School of Physics, Engineering and Technology, University of York, York YO10 5DD, UK}
\affiliation{York-JEOL Nanocentre, University of York, York YO10 5BR, UK}

\author{Demie Kepaptsoglou}%
\affiliation{SuperSTEM, Sci-Tech Daresbury Campus}
\affiliation{School of Physics, Engineering and Technology, University of York, York YO10 5DD, UK}
\author{Quentin M. Ramasse}%
\affiliation{SuperSTEM, Sci-Tech Daresbury Campus}
\affiliation{School of Physics and Astronomy, University of Leeds}
\author{Phil J. Hasnip}%
\affiliation{School of Physics, Engineering and Technology, University of York, York YO10 5DD, UK}%
\author{Vlado K. Lazarov}%
\altaffiliation{vlado.lazarov@york.ac.uk}
\affiliation{School of Physics, Engineering and Technology, University of York, York YO10 5DD, UK}
\affiliation{York-JEOL Nanocentre, University of York, York YO10 5BR, UK}




\date{\today}

\begin{abstract}

We present a methodology based on the calculation of the inelastic scattering from magnons via the spin scattering function in confined geometries such as thin films using a second quantization formalism, for both ferromagnetic and antiferromagnetic materials. The case studies are chosen with an aim to demonstrate the effects of film thickness and crystal orientation on magnon modes, using bcc Fe(100) and NiO with (100) and (111) crystallographic orientations as prototypical systems. Due to the quantization of the quasi-momentum we observe a granularity in the inelastic spectra in the reciprocal space path reflecting the orientation of the thin film. This approach also allows to capture softer modes that appear due to the partial interaction of magnetic moments close to the surface in a thin film geometry, in addition to bulk modes. The softer modes are also affected by crystallographic orientations as illustrated by the different surface-related peaks of NiO magnon density of states at approximately $\sim$ 65 meV for (100) and $\sim$ 42 meV for (111).  Additionally, we explore the role of anisotropy on magnon modes, revealing that introducing anisotropy to both Fe and NiO films increases the overall hardness of the magnon modes. The introduction of a surface anisotropy produces a shift of the surface-related magnon DOS peak to higher energies with increased surface anisotropy, and in some cases leading to surface confined mode. 

\begin{description}
\item[Keywords] Magnonics, magnon dispersion, antiferromagnets, ferromagnets, NiO, bcc Fe
\end{description}
\end{abstract}

\pacs{Valid PACS appear here}
\maketitle


\section{Introduction}

The study of the collective dynamics of magnetic systems has attracted a great deal of attention, due to the promise of using the electron spin to deliver an energetically efficient way to meet the data processing requirements of modern society. This has led to a body of knowledge in a field generally known as spintronics, which encompasses the production, transmission, storage and processing of information using spin-waves \cite{Mahmoud2020,Barman_2021}. Spin waves are dynamic eigen-excitations of magnetically ordered materials, often described in terms of their quanta, or 'magnons', much like photons or phonons are the quanta of light or lattice waves, respectively.
 
The design and optimization of magnonic devices heavily rely on the careful use of specific geometries, particularly emphasizing lower dimensionality, such as thin films. In this context, it becomes imperative to extend current computational methods traditionally applied to the study of 
magnons in bulk materials to take into account effects arising from confined geometries associated with thin films and heterostructures.

As a result, investigations of magnetism in low dimensions serve a dual purpose. First, it offers valuable insights into how the magnetic properties of a solid evolve when transitioning from a three-dimensional (3D) bulk crystal to two, one, or zero-dimensional structures, which often leads to the emergence of unique phenomena not observed in bulk magnets. Second, from a technological standpoint, the incorporation of magnetic materials into modern technologies necessitates their presence in the form of thin films or wires. In fact, low-dimensional magnetic structures play a pivotal role in the field of spintronics \cite{LENK2011107,krawczyk2014review,sheng2021magnonics,petti2022review,wang2021topological}.

Previous computational work on magnons has focused on utilising atomistic spin dynamics \cite{Etz2015}, or on solving the linearised equation of motion that derives from the Heisenberg Hamiltonian \textit{via} a so-called confined ansatz \cite{Cottam2019,Zakeri2016}. Using the latter approach recently a study of magnon confinement effects in low-dimensional magnetic structures was reported ~\cite{Bearisto}, where authors proposed a method for solving the Heisenberg Hamiltonian in the second quantization by transforming the spin operators into bosonic creation and annihilation operators using the Holstein-Primakoff transformation, with calculations performed for a model 1D spin chain. 

In the current work, we extend this approach and provide a generalised method for calculating the effect of confinement in films using the Holstein-Primakoff-transformed Heisenberg Hamiltonian.
This approach allows the application of the machinery of second quantization to thin films with any spin order, including ferromagnets and antiferromagnets, opening a pathway to carrying out calculations in support of the broad range of probing methods used to study magnons in technologically relevant thin film systems. We demonstrate this methodology for ferromagnetic films of bcc Fe(100) and antiferromagnetic NiO with (100) and (111) crystallographic orientations with varying thicknesses, as prototypical systems.

\section{Methods}



We start by expressing the Heisenberg Hamiltonian with a tensorial exchange parameter $J_{ij}$ and a magneto-crystalline anisotropy $K$:

\begin{equation}\label{eq:heisemberg}
    H = -\frac{1}{2}\sum _{ij}\hat{\textbf{S}}_i \cdot J_{ij} \cdot \hat{\textbf{S}}_j -K\sum _{i}(\mathbf{m} \cdot \hat{\textbf{S}}_i)^{2}
\end{equation}

where we are modelling an insulating spin-lattice, with magnetic moments at lattice points labelled by indexes $i$ and $j$ which are represented mathematically by the spin operators, $\textbf{S}_i$ and $\textbf{S}_j$. Here we use a general form for the spin operators, where we take a semi-classical view of the spin operators, justified by the assumption of $S>>1$. Here we use the notation as detailed in \cite{Thz_Spectroscopy}, and expand it to use the exchange interaction as a tensor and treat the system as finite in one direction to study thin films. We can then write our spins in the laboratory frame as:

\begin{equation}
    \textbf{S}_i = S(sin\theta_i cos\phi_i, sin\theta_i sin\phi_i, cos\theta_i)
\end{equation}

We can perform a transformation to the local reference frame of the spin, given by:

\begin{equation}
    \mathcal{R}_{i} =\begin{pmatrix}
cos\theta _{i} cos\phi _{i} & cos\theta_{i} sin\phi_{i} & -sin\theta_{i}\\
-sin\phi _{i} & cos\phi _{i} & 0\\
sin\theta _{i} cos\phi _{i} & sin\theta _{i} sin\phi _{i} & cos\theta _{i}
\end{pmatrix}
\end{equation}

Which allows us to write $\bar{\textbf{S}}_i = \mathcal{R}_{i}\cdot{\textbf{S}}_i$, where $\bar{\textbf{S}}_i$ refers to spins expressed in the local reference frame and ${\textbf{S}}_i$ is the laboratory reference frame. Then, the exchange term of the Heisenberg Hamiltonian, the first term in equation~(\ref{eq:heisemberg}),  can be written as:

\begin{equation}\label{eq:LocalExchange}
\begin{aligned}
H_{ex}= & -\frac{1}{2}\sum _{ij} \hat{\mathbf{S}}_{i} \cdot J_{ij} \cdot \hat{\mathbf{S}}_{j}\\
= & -\frac{1}{2}\sum _{ij}\sum _{\alpha ,\beta ,\gamma }\left(\mathcal{R}_{i}^{-1}\right)_{\alpha \beta }\overline{\mathbf{S}}_{i\beta } (J_{ij})_{\beta\gamma} \left(\mathcal{R}_{j}^{-1}\right)_{\alpha \gamma }\overline{\mathbf{S}}_{j\gamma }\\
= & -\frac{1}{2}\sum _{ij}\sum _{\alpha ,\beta ,\gamma }\overline{\mathbf{S}}_{i\beta }\left(\mathcal{R}_{i}\right)_{\beta \alpha } (J_{ij})_{\beta\gamma}
 \left(\mathcal{R}_{j}^{-1}\right)_{\alpha \gamma }\overline{\mathbf{S}}_{j\gamma }\\
= & -\frac{1}{2}\sum _{ij}\sum _{\alpha ,\beta ,\gamma }\overline{\mathbf{S}}_{i\beta } F_{J}(i,j)_{\beta \gamma }\overline{\mathbf{S}}_{j\gamma }\\
= & -\frac{1}{2}\sum _{ij}\overline{\mathbf{S}}_{i} \cdot F_{J}( i,j) \cdot \overline{\mathbf{S}}_{j}{}
\end{aligned}
\end{equation}

where we have defined:

\begin{equation}
    F_{J}( i,j) = (\mathcal{R}_{i} \cdot J_{ij} \cdot \mathcal{R}_{j}^{-1})
    \label{projection}
\end{equation}
we have used Latin letters to note the lattice sites, and Greek letters for cartesian orientations $x$, $y$ and $z$. Notice that this enables us to define a different orientation for each site of a chosen lattice, allowing studies of ferromagnetic and antiferromagnetic materials as well as different spin textures such as spin canting at the surface.

The second component in the Hamiltonian \eqref{eq:heisemberg} characterizes the magneto-crystalline anisotropy, where $\mathbf{m}$ denotes a unit vector specifying the orientation of the magneto-crystalline anisotropy axis. This vector can be expressed in spherical coordinates as follows:

\begin{equation}\label{eq:MA_direction}
    \mathbf{m}  = ( sin\eta  cos\delta ,sin\eta  sin\delta ,cos\eta )
\end{equation}

where $\eta$ and $\delta$ are the polar and the azimuthal angles, respectively.

The use of a tensorial exchange parameter allows us to model the interaction in a more general way, where the diagonal terms can be recognised as representing the usual constant exchange interaction, while the off-diagonal terms represent the Dzyaloshinskii-Moriya (DM) interaction. Therefore, the tensor terms must follow  $(J_{ij})_{\alpha\beta}=-(J_{ij})_{\beta\alpha}$ for $\alpha\neq\beta$. In other words, the exchange anisotropy manifests itself as an asymmetry in the diagonal terms of the tensorial exchange tensor.

\subsection[Holstein-Primakoff]{Holstein-Primakoff Transformation} \label{Holstein-Primakoff}

In the Heisenberg Hamiltonian (\ref{eq:heisemberg}), the magnetic moments are modelled by spin operators, $\textbf{S}_i$ and $\textbf{S}_j$, which satisfy SU(2) algebra. A usual step to obtain the magnon dispersion is to rewrite the SU(2) operators as bosonic field operators $a_i$ and $a^{\dagger}_i$ \cite{Rezende2020,Thz_Spectroscopy,Nolting2009}, which satisfy the following commutation relations:

\begin{equation}
    [a_i,a^{\dagger}_j] = \delta_{ij}
\end{equation}

\begin{equation}
    [a_i,a_j] = [a^{\dagger}_i,a^{\dagger}_j] = 0
\end{equation}

The corresponding Fourier transformed operators $a_q$ and $a^{\dagger}_q$ correspond to the creation and annihilation of collective oscillations of the magnetic moments, which are interpreted as quasi-particles, the so-called magnons.

The Holstein-Primakoff transformation \cite{HP-Transform} uses this approach to express the spin operators as:

\begin{equation}
\begin{aligned}
S_{i}^{x} & =\frac{\sqrt{2S}}{2}\left( \phi (\hat{n}_i) a_{i} +a_{i}^{\dagger } \phi (\hat{n}_i)\right)
\end{aligned}  
\end{equation}
\begin{equation}
\begin{aligned}
S_{i}^{y} & =\frac{\sqrt{2S}}{2i}\left( \phi (\hat{n}_i) a_{i} -a_{i}^{\dagger } \phi (\hat{n}_i)\right)
\end{aligned}
\end{equation}
\begin{equation}
\begin{aligned}
S_{i}^{z} & =S-\hat{n}_i
\end{aligned}
\end{equation}
where we set $\hbar=1$ by convention, S is the spin quantum number, and $\hat{n}_{i}=a^{\dagger}_ia_i$ is the number operator. We define $\phi(\hat{n}_{i}) = \sqrt{1-\frac{\hat{n}_{i}}{2S}}$ as shorthand notation for convenience. 
 
 Considering that the eigenstates of ${S}^{z}_i$ are $\ket{S,m_s}$ with eigenvalues $m_s=S,S-1,..., -S$, the number operator $\hat{n}_{i}$ has eigenstates $\ket{m'_s}$ where $m'_s = S - m_s$ with eigenvalues $n_i=0,1,2,...,2S$. A physical interpretation of this representation, is that an increase in the number of magnons signifies a decrease in the z-direction projection of the magnetic moment at a particular site.
 
The transformation involves the square root of the operators. Formally, this requires an expansion of the form: 

\begin{equation}
    \phi(\hat{n}_i) = 1 - \frac{\hat{n}_{i}}{4S_i} - \frac{\hat{n}^{2}_{i}}{32S_i^2} - \cdots
\end{equation}

This leads to a Hamiltonian that has an infinite number of terms:

\begin{equation}
    H=\sum_{n=0}^{\infty}H_{n}
\end{equation}

where $H_{n}$ represents a term that involves $n$ operators, e.g., $H_{4}\propto a^{\dagger}_ia_ja^{\dagger}_ka_l$. In this summation scheme, $H_0$ represents the ground state, $H_1$ is zero given the restrictions of asymmetry in the DM interaction, and $H_2$ is the term that carries the spin-wave approximation. Every subsequent term gives rise to multi-magnon interactions. 

For simplicity, we truncate this series using the spin wave approximation, where only the $n=2$ term is used, \textit{i.e.} considering only terms that are quadratic in the operators such that $H_{2}\propto a^{\dagger}_ia_j$. This approximation is justified for temperatures well bellow the magnetic order-disorder transitions temperature, since it leads to removing interactions between magnons, and in cases where $S>>1$ which allow us to neglect quantum fluctuation effects.
\vspace{0.5cm}

\subsection{Fourier Transform of bosonic operators}

Using the results in the previous subsection we can rewrite the Heisenberg Hamiltonian in the second quantization for an arbitrary magnetic moment orientation. Here we are mostly interested in the $H_2$ term of the Hamiltonian which represents the creation of single magnons, or a set of non-interacting magnons. Performing the transformation outlined in \ref{Holstein-Primakoff} with the generalized form given in \eqref{eq:LocalExchange} we get,



\begin{equation}
\begin{aligned}
H_{2} =-\frac{1}{2}\sum _{ij}\Bigl\{ & - S_{i} F_{J}( i,j)_{zz} a_{j}^{\dagger } a_{j} - S_{j} F_{J}( i,j)_{zz} a_{i}^{\dagger } a_{i}\\
+\frac{\sqrt{S_{i} S_{j}}}{2}\Bigl[ & G_{1}( i,j)  a_{i}^{\dagger } a_{j}^{\dagger } +G_{2}^{*}( i,j) a_{i}^{\dagger } a_{j}\\
+ & G_{2}( i,j) a_{i} a_{j}^{\dagger } +G_{1}^{*}( i,j)  a_{i} a_{j}\Bigr]\Bigr\}
\end{aligned}
\label{eq:H2}
\end{equation}



Where we have defined the following short-hand notation, $G_{1}=F_{J}( i,j)_{xx} -iF_{J}( i,j)_{xy} -iF_{J}( i,j)_{yx} -F_{J}( i,j)_{yy}$ and  $G_{2}=F_{J}( i,j)_{xx} +iF_{J}( i,j)_{xy} -iF_{J}( i,j)_{yx} +F_{J}( i,j)_{yy}$, for the terms of the projection matrix.


Next, we take the Fourier transform of the bosonic operators to exploit the periodicity of the system. In the bulk case, the Fourier transform is applied in all three dimensions of the system. In contrast, in the case of thin films, only two dimensions are Fourier transformed,


\begin{equation}
\begin{aligned}
a_{\textbf{q}_{\parallel }}^{( r)} = & \frac{1}{\sqrt{N_{\parallel }}}\sum _{i}^{( r)} e^{-i\textbf{q}_{\parallel } \cdot \textbf{R}_{i}} a_{i}\\
a_{\textbf{q}_{\parallel }}^{( r) \dagger } = & \frac{1}{\sqrt{N_{\parallel }}}\sum _{i}^{( r)} e^{i\textbf{q}_{\parallel } \cdot \textbf{R}_{i}} a_{i}^{\dagger }\\
a_{i} = & \frac{1}{\sqrt{N_{\parallel }}}\sum_{\textbf{q}_{\parallel }} e^{i\textbf{q}_{\parallel } \cdot \textbf{R}_{i}} a_{\textbf{q}_{\parallel }}^{( r)}\\
a_{i}^{\dagger } = & \frac{1}{\sqrt{N_{\parallel }}}\sum_{\textbf{q}_{\parallel }} e^{-i\textbf{q}_{\parallel } \cdot \textbf{R}_{i}} a_{\textbf{q}_{\parallel }}^{( r) \dagger }
\end{aligned}
\label{eq:Fourier}
\end{equation}

The wave vector $\textbf{q}_{\parallel }$ is only defined in directions within the plane of the film, and the normalization constant $N_\parallel$ is the number of magnetic moments in the system, where $r$ labels the sites within the unit cell and $\textbf{R}_{i}$ is the position vector of the ith site. Substituting equation (\ref{eq:Fourier}) within equation (\ref{eq:H2}) and using the following definition for the Kronecker delta,

\begin{equation}
    \sum_{i}e^{i(\mathbf{q}-\mathbf{q'})\cdot \mathbf{r}_{i}} = N\delta_{\mathbf{q}\mathbf{q'}}
\end{equation}
the resulting Hamiltonian matrix is then symmetrized, using the 'spread it around' rule \cite{Thz_Spectroscopy} so that the terms contribute equally to all four quadrants of the resulting matrix.
\begin{widetext}
\begin{equation}
     \begin{array}{l}
H_{2} =\frac{1}{4}\sum {_{q_\parallel}}^{'}\sum _{rs}\sum _{u} z_{u}\Bigl\{\Bigl\{S_{r} F_{J}( r,s)_{zz} a_{q_\parallel}^{( s) \dagger } a_{q_\parallel}^{( s)} + S_{s} F_{J}( r,s)_{zz} a_{q_\parallel}^{( r) \dagger } a_{q_\parallel}^{( r)}\\
-\frac{\sqrt{S_{r} S_{s}}}{2}\Bigl[ G_{1}( r,s)  \Gamma _{rs}^{*( u)}( q_\parallel) a_{q_\parallel}^{( r) \dagger } a_{-q_\parallel}^{( s) \dagger } +G_{1}^{*}( r,s) \Gamma _{rs}^{( u)}( q_\parallel) a_{-q_\parallel}^{( r)} a_{q_\parallel}^{( s)} +G_{2}( r,s) \Gamma _{rs}^{( u)}( q_\parallel) a_{q_\parallel}^{( r) \dagger } a_{q_\parallel}^{( s)} +G_{2}^{*}( r,s) \Gamma _{rs}^{*( u)}( q_\parallel) a_{q}^{( r) \dagger } a_{q_\parallel}^{( s)}\Bigr]\Bigr\}\\
                              \hspace{3.75cm}  +\Bigl\{S_{r} F_{J}( r,s)_{zz} a_{-q_\parallel}^{( s)} a_{-q_\parallel}^{( s) \dagger }  +S_{s} F_{J}( r,s)_{zz} a_{-q_\parallel}^{( r)} a_{-q_\parallel}^{( r) \dagger }\\
       -\frac{\sqrt{S_{r} S_{s}}}{2}\Bigl[ G_{1}( r,s) \Gamma _{rs}^{*( u)}( q_\parallel) a_{q_\parallel}^{( r) \dagger } a_{-q_\parallel}^{( s) \dagger } +G_{1}^{*}( r,s)  \Gamma _{rs}^{( u)}( q_\parallel) a_{-q_\parallel}^{( r)} a_{q_\parallel}^{( s)} +G_{2}( r,s) \Gamma _{rs}^{( u)}( q_\parallel) a_{-q_\parallel}^{( r)} a_{-q_\parallel}^{( s) \dagger } +G_{2}^{*}( r,s) \Gamma _{rs}^{*( u)}( q_\parallel) a_{-q_\parallel}^{( r)} a_{-q_\parallel}^{( s) \dagger }\Bigr]\Bigr\}\Bigr\}
\end{array}
\label{eq:H_2_fourier}
\end{equation}
\end{widetext}

where $\Gamma_{rs}^{( u)} = (1/z_{u})\sum_{\textbf{d}_u} e^{-i\textbf{q}_\parallel \cdot \textbf{d}_u} $ and $\textbf{d}_u$ is one of the $z_u$ different u'th nearest-neighbour distance vectors, while $r$ and $s$ label the different magnetic moments in the unitcell. The magneto-crystalline term can also be rewritten using similar steps. Since every term in equation (\ref{eq:H_2_fourier}) has a pair of creation and annihilation operators, we can write it as a matrix multiplication of a matrix with the numerical values of the system dependent parameters $F_{J}( r,s)$ and $\Gamma _{rs}^{( u)}( q_\parallel)$, with two vectors composed of the creation and annihilation operators. We can then write \eqref{eq:H_2_fourier} in a compact form as follows:

\begin{equation}
    H_2=v^{\dagger}_{q_\parallel}\cdot \textbf{L} \cdot v_{q_\parallel}
\end{equation}
where we defined:

\begin{equation}
v^{\dagger} = (a^{(1)\dagger}_{q_\parallel},...,  a^{(M)\dagger}_{q_\parallel}| a^{(1)}_{-q_\parallel},...,a^{(M)}_{-q_\parallel})
\label{eq:RealBasis}
\end{equation}
which has the commutation relation:

\begin{equation}
\begin{aligned}
\left[ v_{q_\parallel} ,v_{q_\parallel'}^{\dagger }\right] = & N\delta _{q_\parallel,q_\parallel'}\\
[ v_{q_\parallel} ,v_{q_\parallel'}] =\left[ v_{q_\parallel}^{\dagger } ,v_{q_\parallel'}^{\dagger }\right] = & 0
\end{aligned}
\end{equation}
where:
\begin{equation}
    N=\begin{pmatrix}
    I & 0\\
    0 & -I
\end{pmatrix}
\end{equation}
with $I$ being the identity matrix with rank half the length of $v^{\dagger}$.

Taking into account these commutation relations we obtain the equation of motion for $v_{q}$,

\begin{equation}
    i\frac{dv_{q_\parallel}}{dt}=-[H_2,v_{q_\parallel}]=\mathcal{L}(q_\parallel) \cdot v_{q_\parallel}
\end{equation}
where $\mathcal{L}(q_\parallel)=L(q_\parallel)\cdot N$ where we used $\hbar=1$.

Assuming that the unit cell has M magnetic moments, the matrix $\mathcal{L}(q_\parallel)$ will be 2M-dimensional, with real eigenvalues when the structure is stable, such that $\varepsilon_n(q_\parallel)=\omega_n(q_\parallel)/2 \geq 0$ for $n=1,\cdots,M $ and $\varepsilon_n(q_\parallel)=-\omega_{n}(q_{\parallel})/2 \leq 0$ for $n=M+1,\cdots,2M$, with $\hbar=1$ i.e. the eigenvalues of $\mathcal{L}$ are related to the eigenenergies of the magnon modes that are allowed in the system for each $q$. There will be M positive and M negative eigenvalues due to particle-hole symmetry.

We diagonalize $L(q)$ with the unitary transformation $L'(q)=UL'(q)U^{\dagger}$, where $U^{\dagger}$ is a matrix which  columns are the eigenvectors of $L$, this allow us to write:

\begin{equation}
    H_2=\vec{v}^{\dagger}U^{\dagger}U \textbf{L}U^{\dagger}U \vec{v}=w^{\dagger}L'w
\end{equation}
having defined $w^{\dagger} = \vec{v}^{\dagger}U^{\dagger}$ given by:

\begin{equation}
w^{\dagger} = (\alpha^{(1)\dagger}_{q_\parallel},...,  \alpha^{(M)\dagger}_{q_\parallel}| \alpha^{(1)}_{-q_\parallel},...,\alpha^{(M)}_{-q_\parallel})
\label{eq:ProjectedRealBasis}
\end{equation}

Comparing with equation \eqref{eq:RealBasis} we can define:

\begin{equation}
    \alpha^{(r)\dagger}_{q_\parallel} =\sum _{n=1}^{N}\left( U_{r,n}^{\dagger } a^{(n)}_{q_\parallel}  + U_{r,n+N}^{\dagger } a^{(n)\dagger }_{-q_\parallel} \right)
    \label{eq:proj1_recip}
\end{equation}
\begin{equation}
    \alpha^{(r)\dagger}_{-q_\parallel} =\sum_{n=1}^{N} \left( U_{r+N,n}^{\dagger } a_{q_\parallel}^{(n)}  + U_{r+N,n+N}^{\dagger} a^{(n)\dagger}_{-q_\parallel} \right)
    \label{eq:proj2_recip}
\end{equation}
while in real space we have:

\begin{equation}
    a_{j} =\sum _{n=1}^{N}\left( U_{j,n}^{\dagger } \alpha _{n}  + U_{j,n+N}^{\dagger } \alpha _{n}^{\dagger }\right)
    \label{eq:proj1_real}
\end{equation}
\begin{equation}
    a_{j}^{\dagger } =\sum _{n=1}^{N}\left( U_{j+N,n}^{\dagger } \alpha _{n}  + U_{j+N,n+N}^{\dagger } \alpha _{n}^{\dagger }\right)
    \label{eq:proj2_real}
\end{equation}

Expanding $H_2$ in the new diagonal basis with the eigenfrequencies obtained from the diagonalization of  $\mathcal{L}(q_\parallel)$ we can write,

\begin{equation}
    H_2 = \sum_{n=1}^{M}\sum_{q_\parallel} \omega_n(q_\parallel) \left[ \alpha^{(n)\dagger}_{q_\parallel}\alpha^{(n)}_{q_\parallel} + \frac{1}{2}\right]
    \label{eq:OscilationHamilt}
\end{equation}

This approach allows the calculation of observables such as the spin scattering function,  magnetization, etc., for both bulk or thin films, by projecting the second quantized spin operators onto the diagonalized basis of the Hamiltonian $\alpha _{n}$ and $\alpha _{n}^{\dagger }$, using equations \eqref{eq:proj1_real} and \eqref{eq:proj2_real}. 


\subsection{Spin Scattering function}

The spin scattering function is a proxy for neutron scattering measurements also sometimes called the dynamic structure factor. It is given by the space and time Fourier transform of the time-dependent spin-spin correlation function:

\begin{equation}
    S_{\alpha \beta }(\mathbf{q} ,\omega ) =\frac{1}{2\pi N}\sum _{j,k}\int dte^{-i\omega t} e^{-i\mathbf{q} \cdot (\mathbf{r}_{j} -\mathbf{r}_{k})} \langle s_{j}^{\alpha }( 0) s_{k}^{\beta }( t) \rangle_T
    \label{eq:spinspinfunction}
\end{equation}

Here we are noting $N$ as the total number of spins in the lattice. Once more, we employ Latin letters to designate lattice sites, while Greek letters are utilized for denoting Cartesian directions. This way the spin operator $s_{j}^{\alpha }$ represents the $\alpha$ component of a spin located at position $\mathbf{r}_{j}$. The $\langle \cdot \rangle_T$ denotes the quantum and thermal average at temperature $T$.

In this section, we will connect the spin scattering function and the eigenvalues and eigenvectors of $\mathcal{L}$, which will give us the frequencies and their spectral weights for the magnons in the system, respectively.

 The spin operators in real space can be expressed as:

\begin{equation}
s_{j}^{\alpha }( t) =\sqrt{\frac{S}{2}}\left\{V_{j\alpha }^{-} a_{j}( t) +V_{j\alpha }^{+} a_{j}^{\dagger }( t)\right\}
\label{eq:transformSimple}
\end{equation}

noting that $a_j(t)=e^{-i\omega_n t}a_j$, and with:

\begin{equation}
    V_{j\alpha }^{\pm} = (\mathcal{R}_{j})_{x\alpha} \pm i(\mathcal{R}_{j})_{y\alpha}
\end{equation}
Substituting \eqref{eq:transformSimple}, \eqref{eq:proj1_real} and \eqref{eq:proj2_real}, in the scattering function \eqref{eq:spinspinfunction} and noting that $\langle \alpha _{n}^{\dagger } \alpha _{n'}^{\dagger } \rangle _{T} =\langle \alpha _{n} \alpha _{n'} \rangle _{T} =0$ and $\langle \alpha _{n}^{\dagger } \alpha _{n'} \rangle _{T} =n_{B}( \omega _{n}) \delta_{nn'}$, where $n_{B}$ is the Bose-Einstein distribution. We perform the Fourier transform in the scattering function and focus only on the positive energies of the spectrum, we get,

\begin{widetext}
\begin{equation}
S_{\alpha \beta }(\mathbf{q} ,\omega ) =\frac{1}{2N_{\parallel }}\sum _{n=1}^{M}\sum _{r,s}\frac{\sqrt{S_{r} S_{s}}}{2} e^{-i\mathbf{q}_{\perp } \cdot ( \mathbf{r}_{r} -\mathbf{r}_{s})}\left( W_{r}^{( n)}\right)_{\alpha }\left( W_{s}^{( n)}\right)_{\beta }[ 1+n_{B}] \delta ( \omega -\omega _{n})
\label{eq:ScatteringFunction_final}
\end{equation}
 \end{widetext}


 where we have defined:

\begin{equation}
    \left( W_{r}^{( n)}\right)_{\alpha } =\left( V_{r,\alpha }^{-} U_{r,n}^{\dagger } +V_{r,\alpha }^{+} U_{r+N,n}^{\dagger }\right) 
\end{equation}

\begin{equation}
     \left( W_{s}^{( n)}\right)_{\beta } =\left( V_{s,\beta }^{-} U_{s,n+N}^{\dagger } +V_{s,\beta }^{+} U_{s+N,n+N}^{\dagger }\right)
\end{equation}

To account for finite instrument resolution we replace the $\delta ( \omega -\omega _{n})$ by a Gaussian broadening given by:

\begin{equation}
    \delta ( \omega -\omega _{n}) = \frac{1}{\sqrt{2\pi \Delta ^{2}}} e^{-\frac{( \omega -\omega _{n})^{2}}{2\Delta ^{2}}}
    \label{eq:Gaussian}
\end{equation}

It is important to emphasise that with this method we can evaluate thin films and heterostructures, and also use the parameters such as $J_{ij}$ and anisotropies $K$ for each different layer and between different layers, as required by the physics of the system.

For colinear spins, the calculation of  $S_{\alpha\beta}$, where $\alpha$ and $\beta$ are the Cartesian directions \textit{x}, \textit{y} and \textit{z}, aligned with the z-axis, we can assume that $S_{xx}(q,\omega) = S_{yy}(q,\omega) \neq 0$ while $S_{\alpha\beta}(q,\omega) = S_{zz}(q,\omega) = 0$ for $\alpha \neq \beta$. 

The spin scattering function gives a relative measurement of the amount the magnetic moment changes in a particular direction, \textit{i.e.}, if all magnetic moments in the system are pointing in the \textit{z}-direction and start to precess, only the magnetic moment projections in the \textit{x} and \textit{y}-directions change, which leads to a non-zero value for the $S_{xx}(q,\omega)$ and $S_{yy}(q,\omega)$ and zero value for $S_{zz}(q,\omega)$, assuming the spin-wave approximation.
\\
\subsection{Magnon density of states}

To compute the magnon density of states (DOS), we employ a methodology akin to that utilized in the derivation linking the velocity correlation function to the vibrational density of states for phonons \cite{VelocityPhonon1,VelocityPhonon2}. Analogously, we extend this approach to the spin-spin correlation function. Employing the Holstein-Primakoff transformation, and the change of basis given in equations \eqref{eq:transformSimple}, \eqref{eq:proj1_real} and \eqref{eq:proj2_real} in the time-dependent form $\alpha_n(t)=\alpha_ne^{-i\omega_{n}t}$ were $\omega_{n}$ are the eigenfrequencies of the magnons in the system, we can derive a formula for the density of states given by:


\begin{widetext}
\begin{equation}
\rho (\omega ) \propto  \sum_{n=1}^{M}\sum_{q}\sum_{r,s} e^{-i\mathbf{q}_{\perp } \cdot (\mathbf{r}_{r} -\mathbf{r}_{s} )}\left( W_{r}^{(n)}\right)_{\alpha }\left( W_{s}^{(n)}\right)_{\alpha } \delta (\omega -\omega _{n} (q))
\label{eq:DOS11}
\end{equation}
\end{widetext}
where we used Einstein's summation convention for the Cartesian orientations labelled by $\alpha$ and the constants in front were omitted. This approach ensures uniformity in the density of states (DOS), irrespective of the number of magnetic moments within the unit cell representing the system. Each magnetic moment in the unit cell contributes to a distinct mode in the magnon dispersion, with no guarantee of degeneracy. By deriving the DOS through the spin-spin correlation function, distinct weights are assigned to each mode, with a preference for those associated with the primitive cell. This weighting scheme ensures a consistent solution, rendering the final DOS independent of the chosen unit cell for calculation purposes.

\section{Results}\label{Results}

In this section, we will provide examples to show the physics that is captured by the method outlined above. We will go over three different cases of thin films, ferromagnetic bcc Fe (100), antiferromagnetic NiO(100) and NiO(111). We will evaluate numerically the spin scattering function using equation \eqref{eq:ScatteringFunction_final}, and the magnon DOS by calculating the contribution to the spin scattering function of all modes in a grid in reciprocal space.

\subsection{bcc Fe(100)}

We start with the prototype ferromagnetic material \textit{bcc} Fe, a well-known and broadly studied system \cite{Oppeneer1998,Pajda2001,Etz_2015,Durhuus_2023}.

The calculations performed in this section have used the $J_{ij}$ parameters proposed in \cite{Oppeneer1998}, where 25 nearest neighbours are taken into account, and the magnetic moment given in \cite{Kittel2004}.

\begin{figure*}[h!]
    \centering 
\begin{subfigure}{0.316455696\textwidth}
    \raggedright
    (a)
    \includegraphics[width=\linewidth]{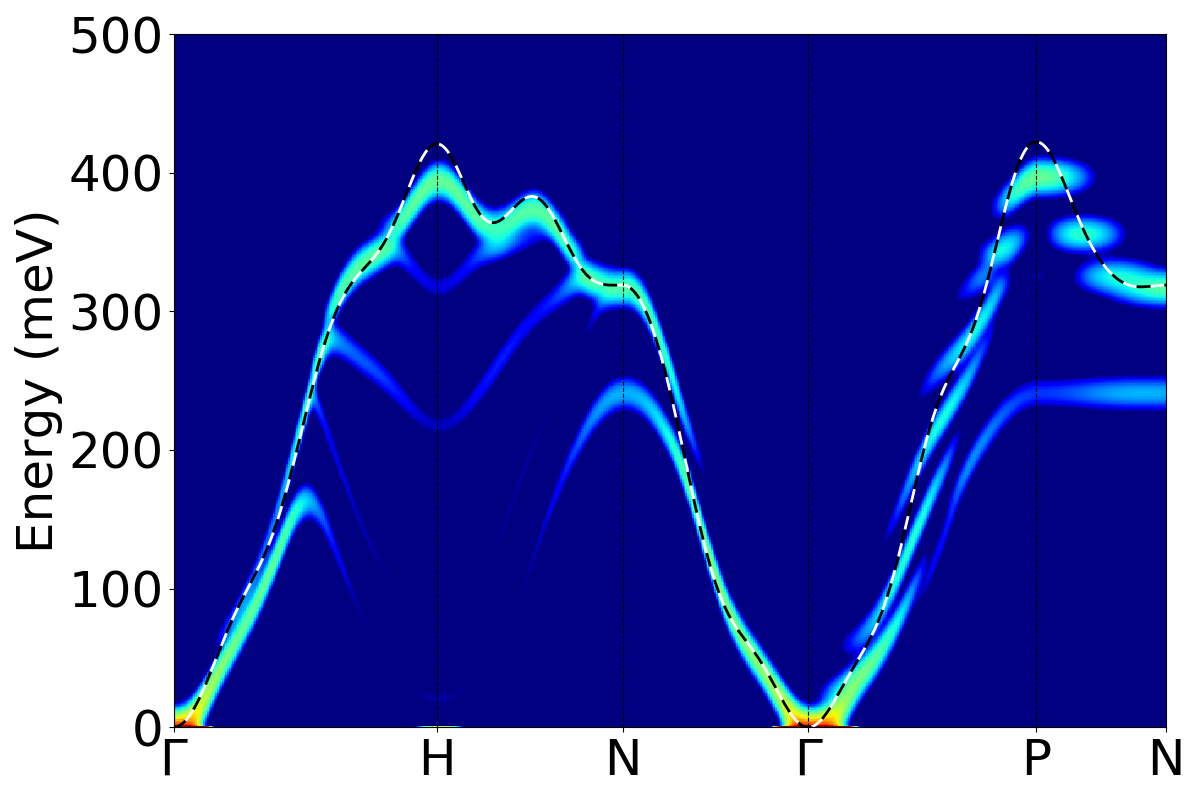}

  \label{fig:1}
\end{subfigure}\hfil 
\begin{subfigure}{0.316455696\textwidth}
    \raggedright
    (b)
    \includegraphics[width=\linewidth]{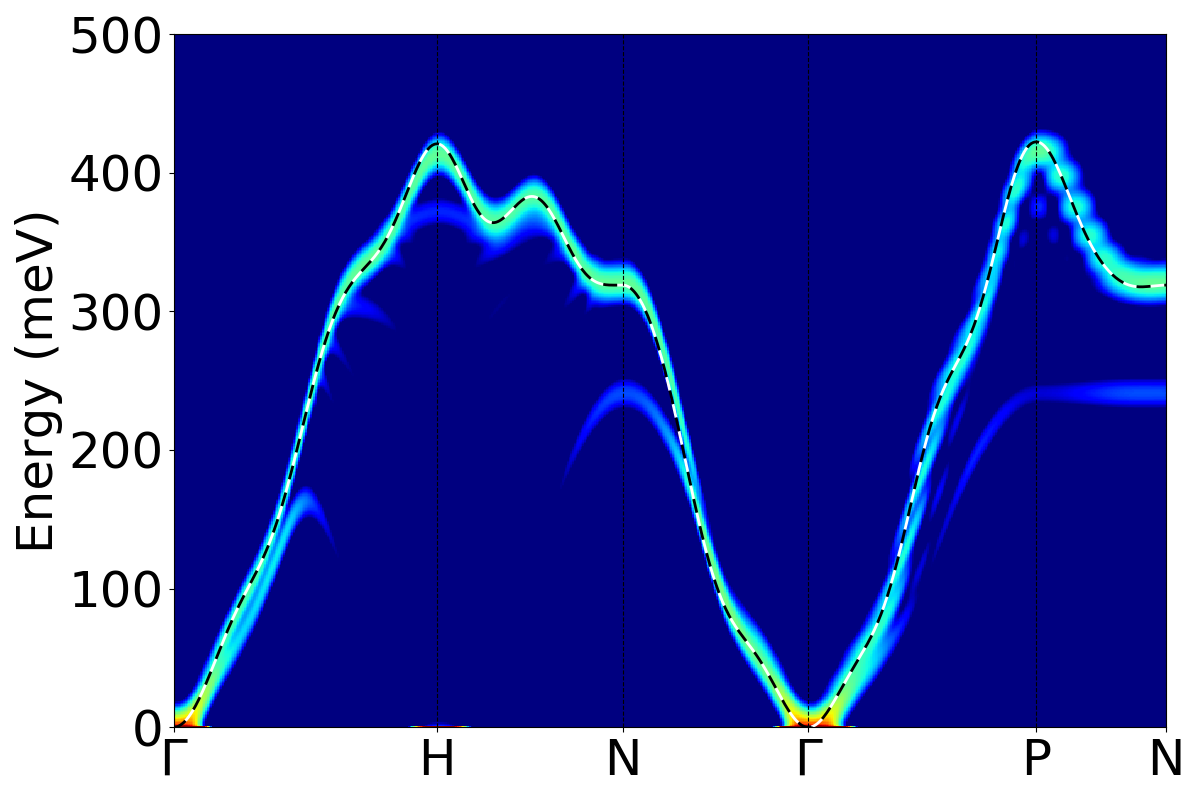}
    
  \label{fig:2}
\end{subfigure}\hfil 
\begin{subfigure}{0.367088608\textwidth}
    \raggedright
    (c)
  \includegraphics[width=\linewidth]{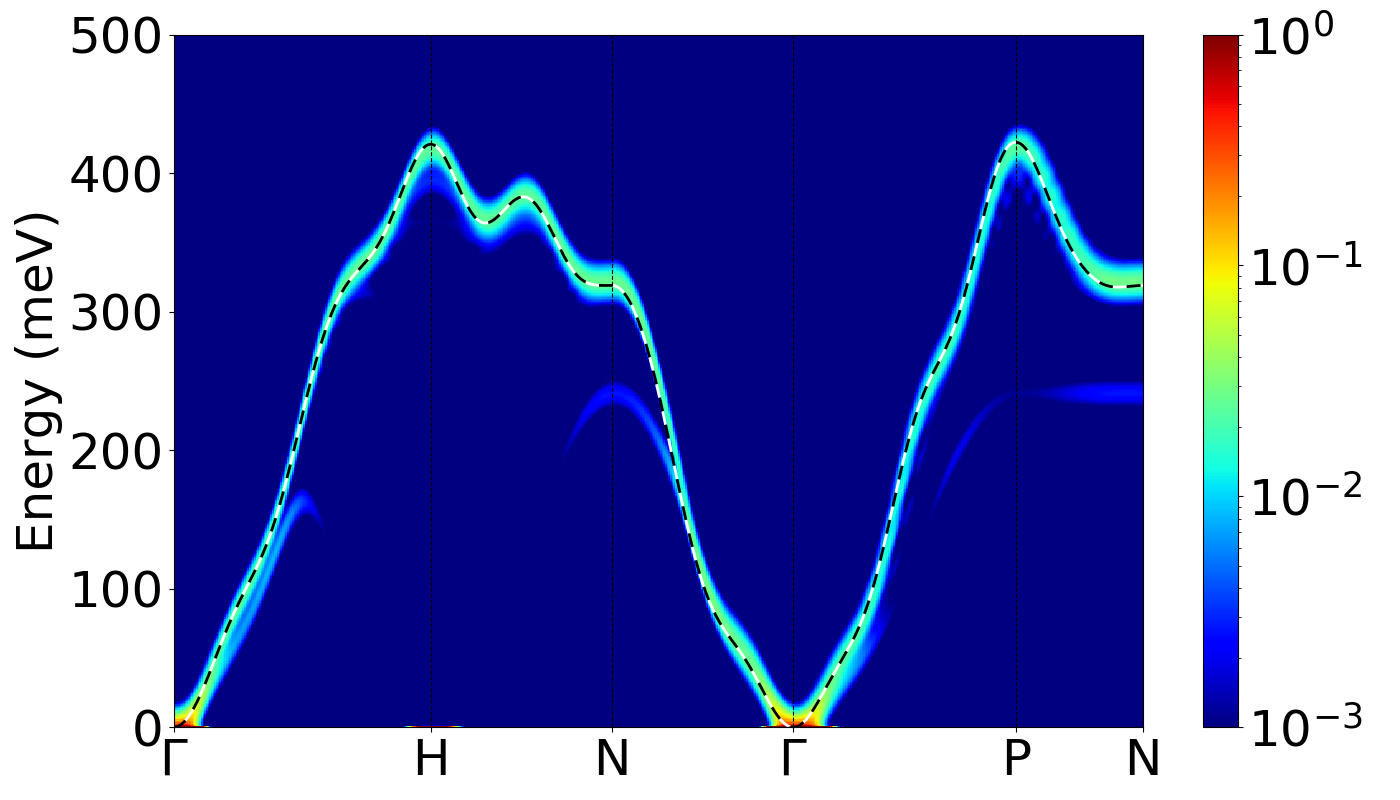}

  \label{fig:3}
\end{subfigure}

\medskip
\begin{subfigure}{0.315955766\textwidth}
    \raggedright
    (d)
  \includegraphics[width=\linewidth]{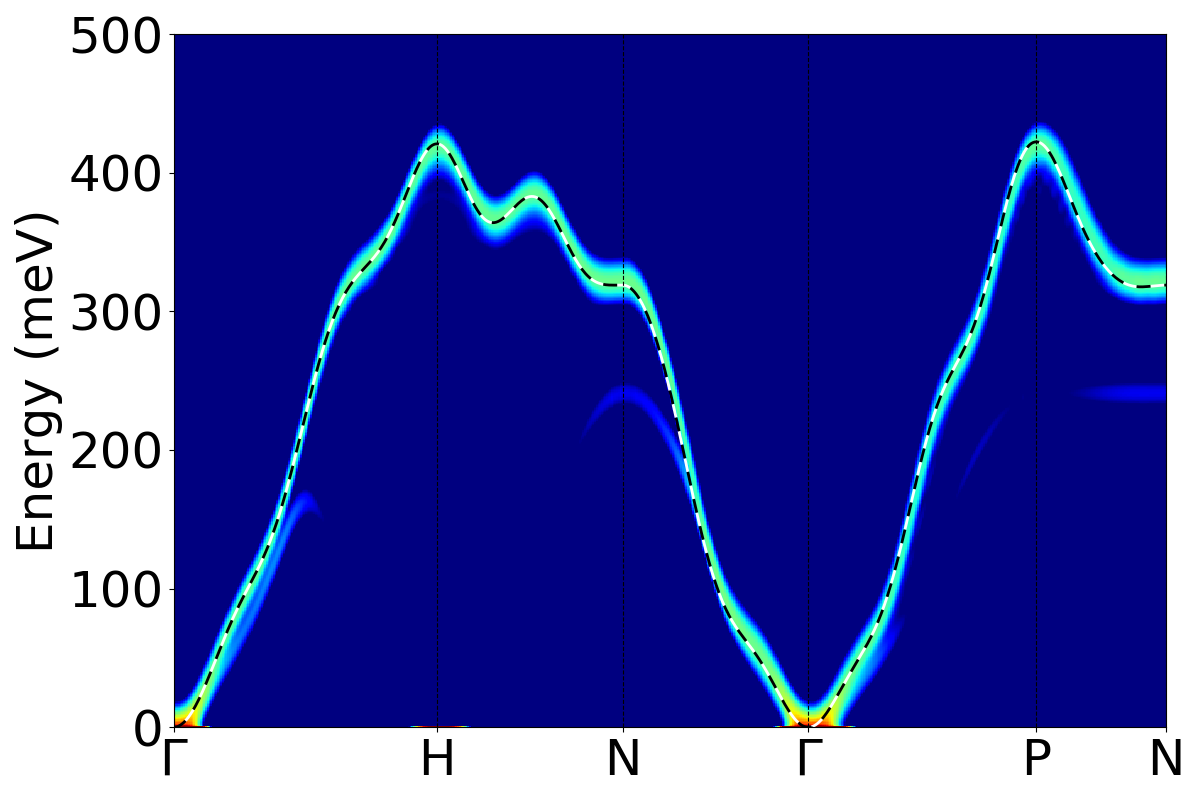}

  \label{fig:4}
\end{subfigure}\hfil 
\begin{subfigure}{0.315955766\textwidth}
    \raggedright
    (e)
  \includegraphics[width=\linewidth]{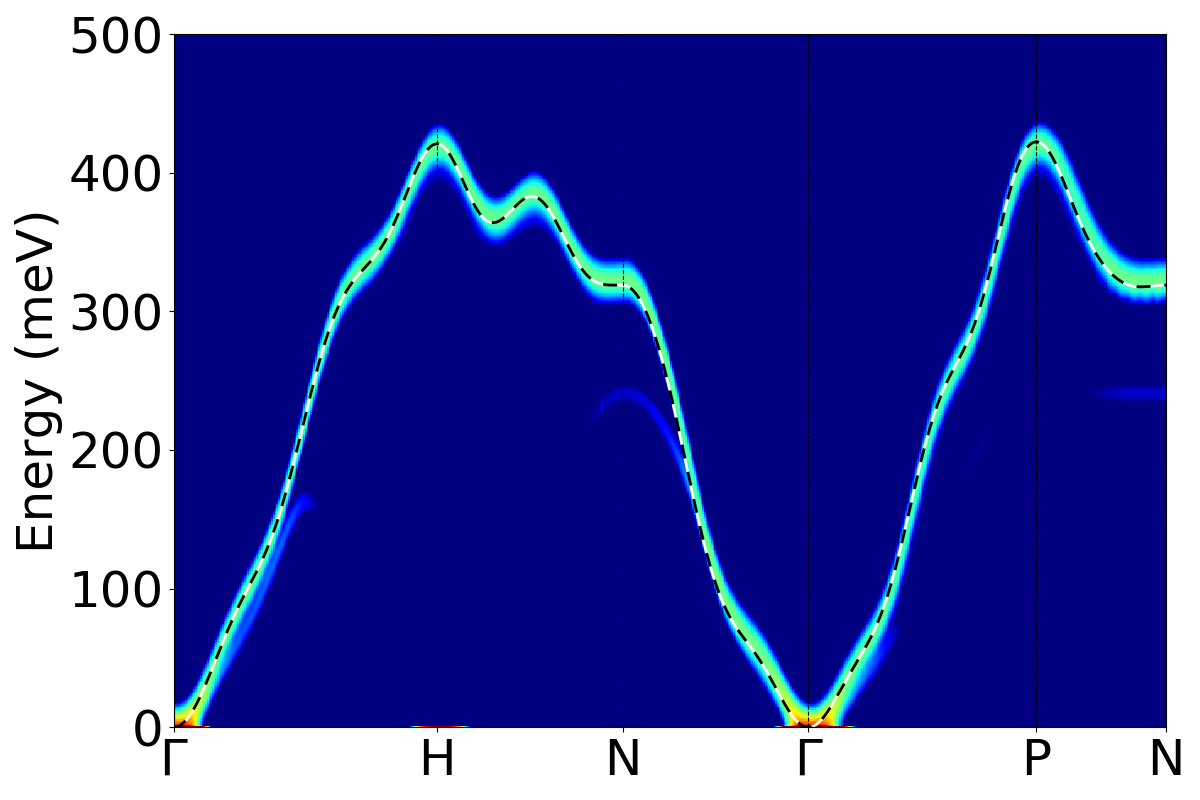}

  \label{fig:5}
\end{subfigure}\hfil 
\begin{subfigure}{0.368088468\textwidth}
    \raggedright
    (f)
  \includegraphics[width=\linewidth]{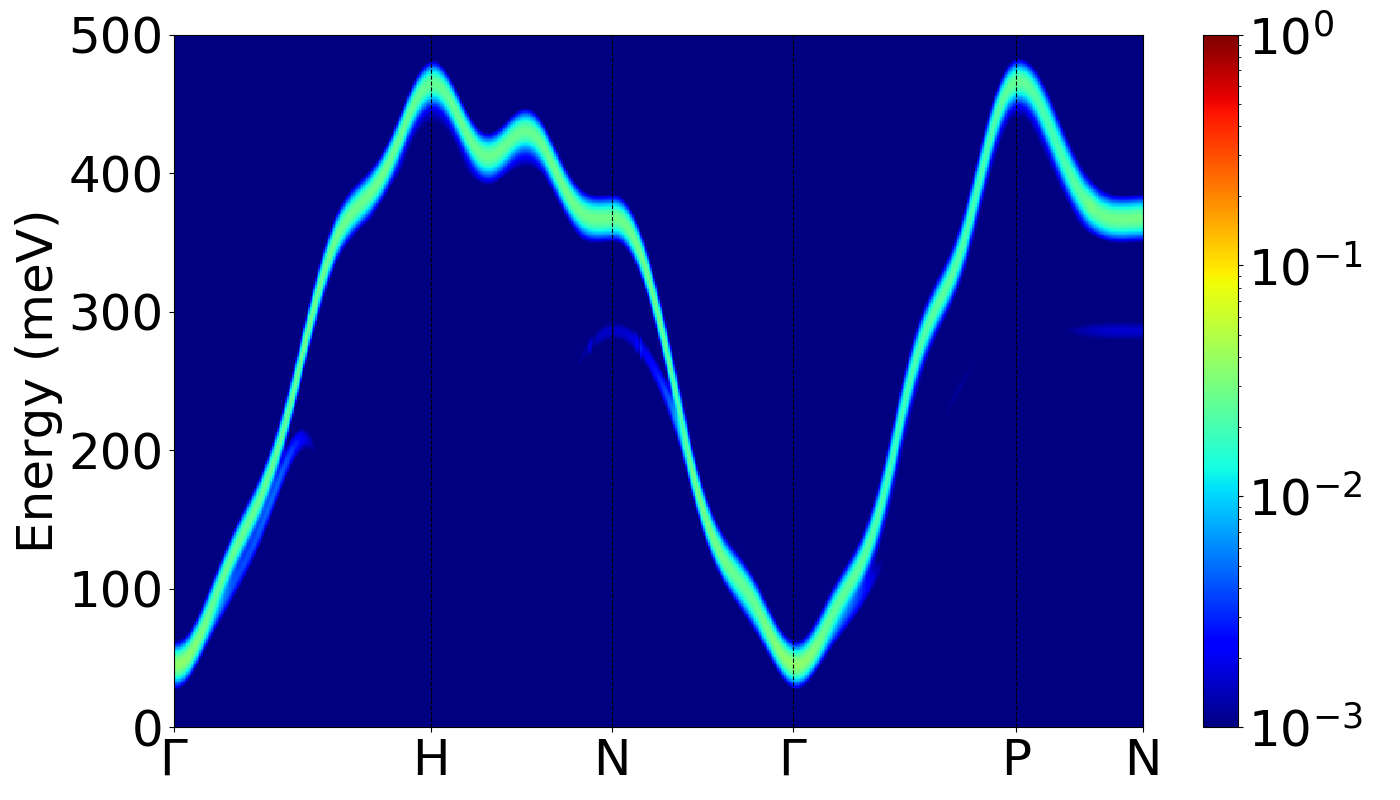} 

  \label{fig:6}
\end{subfigure} 
\caption{Spin scattering function of bcc Fe thin films with a)10, b)20, c)30, d)40, e)50 monolayers f) 50 monolayers with an added 20 meV of anisotropy in the direction of the magnetic moments, with 25 nearest neighbours and a temperature of 300K. For comparison, the bulk dispersion is shown as a black/white line. }
\label{fig:Febcc1}
\end{figure*}

In figure \ref{fig:Febcc1} we show the $S_{xx}$ for 300K, for a thin film of bcc Fe in (001) orientation. In other words, the sample is assumed to be infinite in the $(x,y)$ plane, and confined to a limited number of monolayers in the $z$ direction. As bcc Fe is a cubic system, $S_{xx}$ is the same as $S_{yy}$. The spin scattering functions are evaluated for thin films with sizes ranging from 10 to 50 monolayers. In addition, we show the bulk magnon dispersion, calculated using our formulation but for an infinite model in 3D, as a dashed black/white line. The effect of confinement is clearly observed by the granularity in the spin scattering function plot arising from quasi-momentum quantization resulting from the finite nature of this direction, as shown along the path traversing the finite z-direction ($\Gamma - P - N$). We observe that, since the trajectory deviates from the $k_z$ direction, which is the direction of confinement in our thin film geometry, the quantization does not yield symmetrical shapes for the granular features. Instead, the resulting features exhibit distinct shapes attributable to the trajectories in reciprocal space that are non-parallel to the finite direction. As we increase the number of monolayers, these granular features become increasingly densely packed, ultimately converging towards a continuous line in the limit of a bulk solid. The thin film calculations therefore tend towards the bulk when the system size increases, as expected in the limit of a large number of monolayers. We note that the calculated bulk spin scattering function using our approach agrees well with previous calculations of the magnon dispersion diagram and accompanying experiments for bulk \textit{bcc} Fe \cite{Oppeneer1998,Etz_2015}. In particular, our results appear to capture the onset of the Kohn anomalies in the path between $\Gamma - H$ and between $H - N$, alebit less prominently than in references \cite{Oppeneer1998,Etz_2015}, which can be explained due to the fact that we have employed parameters representing next nearest neighbours up to a distance of 5 times the lattice constant $a$, whereas prior research had considered parameters extending up to 7$a$ as discussed in \cite{Oppeneer1998,Etz_2015}.

The treatment of the Heisenberg Hamiltonian as presented in the methods captures the fact that the top and bottom layers (\textit{i.e.}, the surfaces) have fewer neighbours than bulk-like layers, resulting in a reduction of interactions, which leads to the appearance of softer modes that are less intense and decrease in intensity as the system increases in size, see Fig. \ref{fig:Febcc1}. Furthermore, in Fig. \ref{fig:Febcc1}(f) we show that the addition of $K=20$  meV of magneto-crystaline anisotropy in the direction of the magnetic moments in the 50 monolayers case, results in a rigid shift in energy for all the modes to higher energies. 

\begin{figure}[!h]
    \centering
    \includegraphics[scale=0.18]{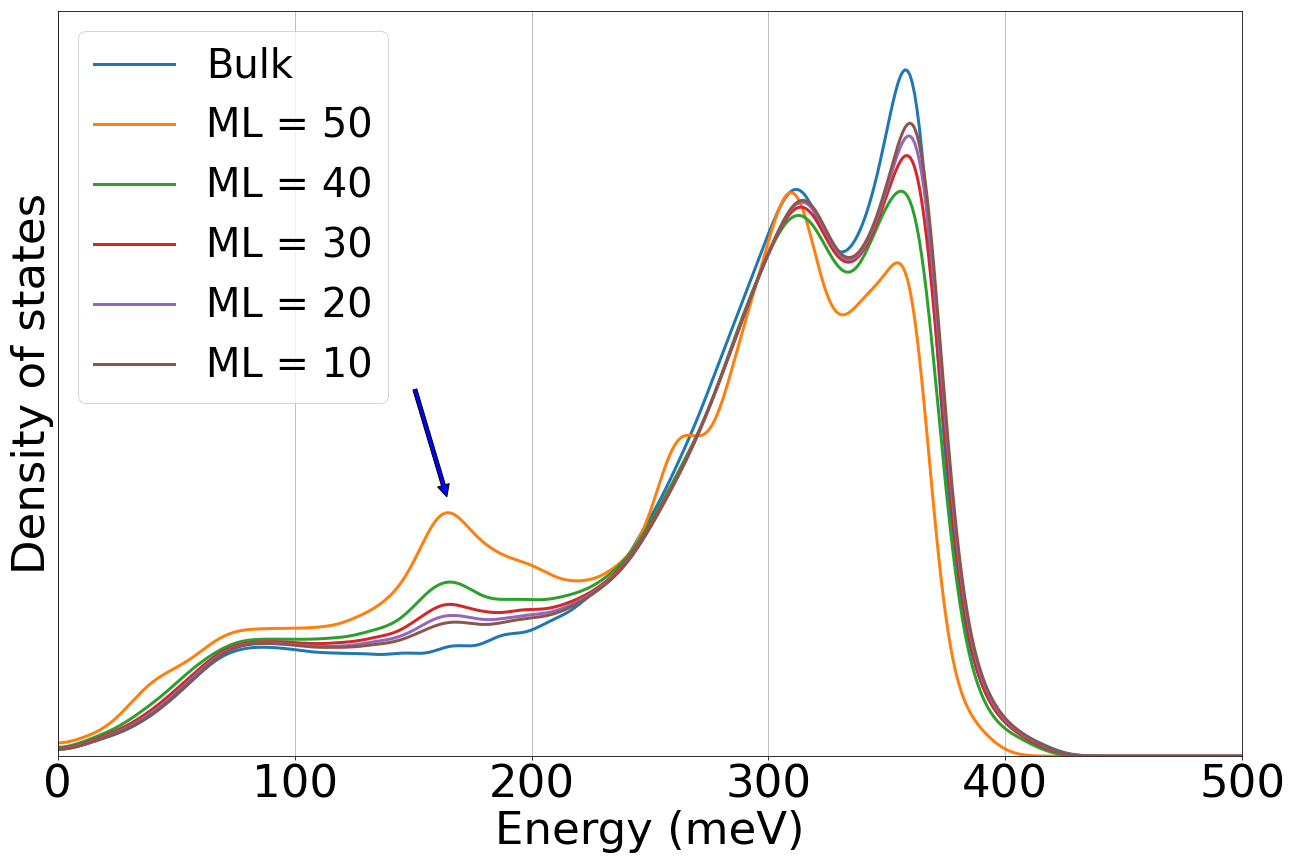}
    \caption{Density of states of Fe bcc comparing Bulk with varying sizes of thin films with 25 nearest neighbours and a temperature of 300K. }
    \label{fig:FebccML}
\end{figure}

The existence of the additional softer surface-related modes can be clearly seen in the magnon DOS calculated using \eqref{eq:DOS11}, as shown in Fig. \ref{fig:FebccML}, for a thin film of 10 monolayers. As highlighted by a blue arrow we can see the appearance of a peak around 180 meV, which is not present in the bulk DOS.

\begin{figure}[!h]
    \centering
    \includegraphics[scale=0.18]{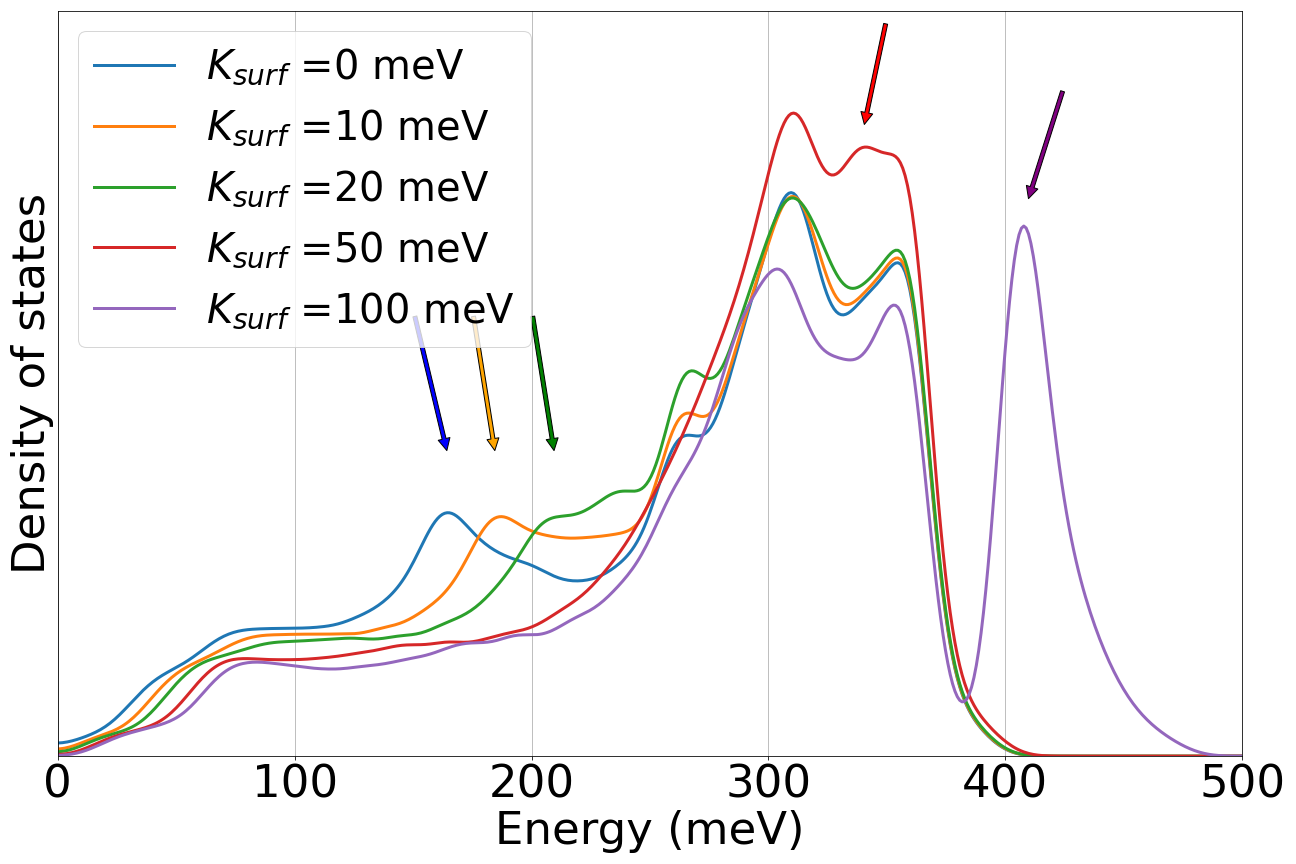}
    \caption{Density of states of bcc Fe thin film with size 10 monolayers, for varying intensities of $K_{surf}$ with 25 nearest neighbours and a temperature of 300K.}
    \label{fig:FebccSurfAni}
\end{figure}

This confined peak in the DOS is dependent on the surface properties. This is illustrated in Fig. \ref{fig:FebccSurfAni}, where on the top and bottom surfaces of the 10 monolayer Fe film we added an artificial surface-only anisotropy ($K_{surf}$). As $K_{surf}$ increases, the confined DOS peak shifts to higher energies, and eventually becomes localized at energies above the bulk dispersion. 

\subsection{NiO(100) and NiO(111)}

Next, we consider NiO to demonstrate the applicability of our method to antiferromagnetic thin films. The flexible ability to set the magnetic moments modulus and directions inside the unit cell allows us to study, both colinear and non-colinear systems as well as the interfaces between them.

We used parameters from inelastic neutron scattering experiments that suggest that the first-neighbour ferromagnetic exchange interaction parameters are $J_{1,p}= -1.39$ meV and $J_{1,ap}= -1.35$ meV, where $J_{1,p}$ is the interaction between parallel first neighbours and $J_{1,ap}$ is the interaction between anti-parallel first neighbours \cite{NiO_J1, Hutchings1972}. Similarly, the second-neighbour antiferromagnetic exchange interaction was determined to be $J_2=19.01$  meV. The small difference between $J_{1,p}$ and $J_{1,ap}$ was attributed to lattice distortion, as previously pointed out \cite{NiO_J1, Hutchings1972}.

\begin{figure*}[h!]
    \centering 
\begin{subfigure}{0.315955766\textwidth}
    \raggedright
    (a)
  \includegraphics[width=\linewidth]{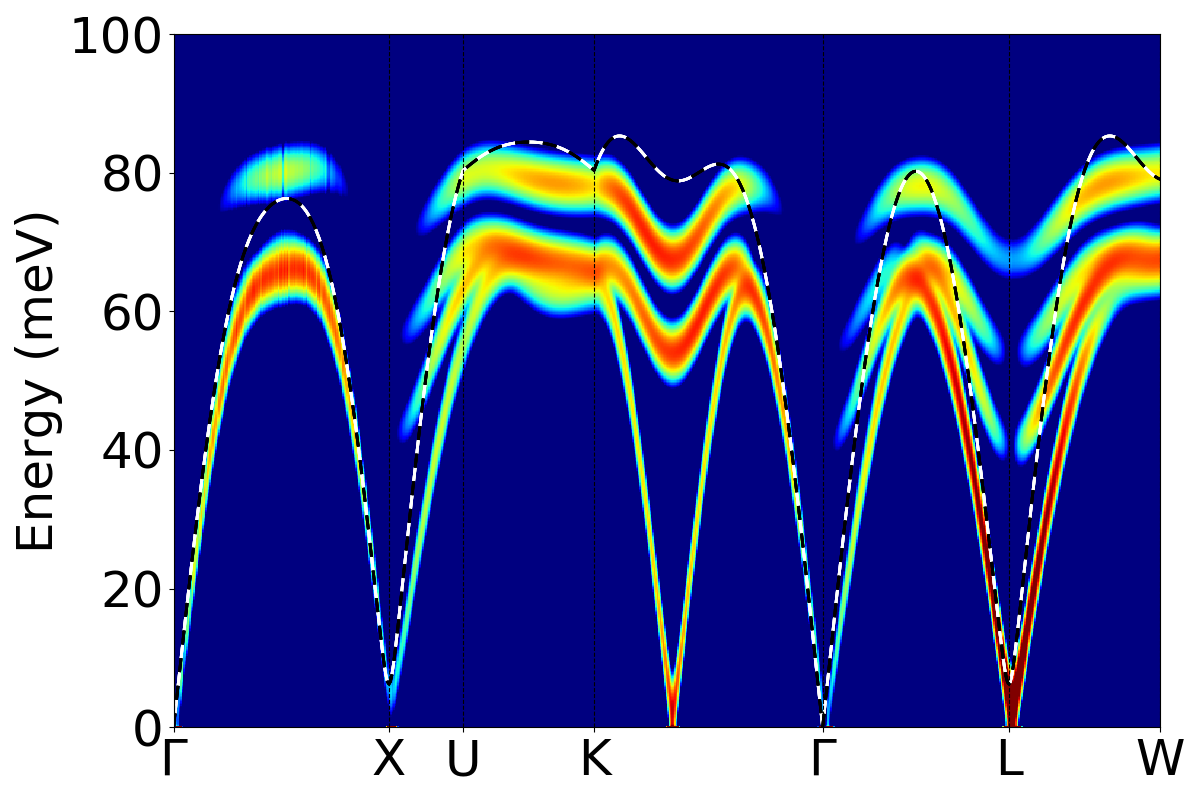}

  \label{fig:NiO_1}
\end{subfigure}\hfil 
\begin{subfigure}{0.315955766\textwidth}
    \raggedright
    (b)
  \includegraphics[width=\linewidth]{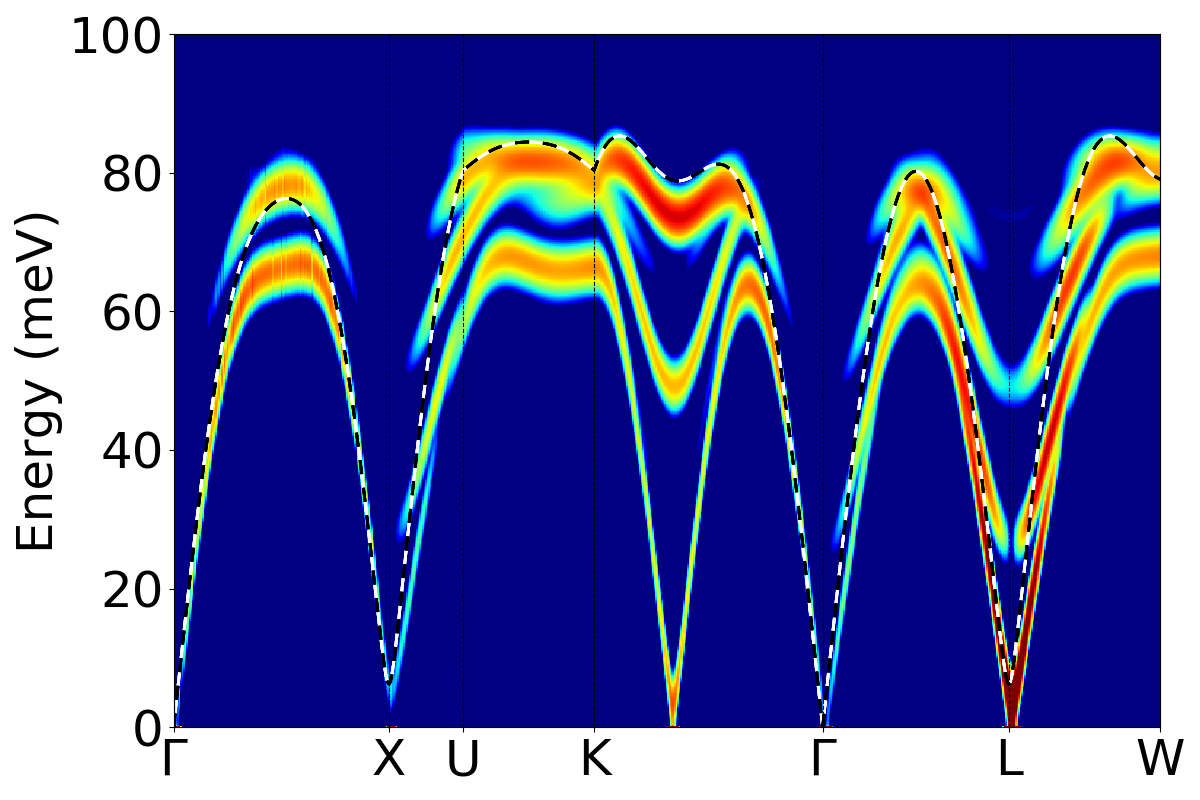}

  \label{fig:NiO_2}
\end{subfigure}\hfil 
\begin{subfigure}{0.368088468\textwidth}
    \raggedright
    (c)
  \includegraphics[width=\linewidth]{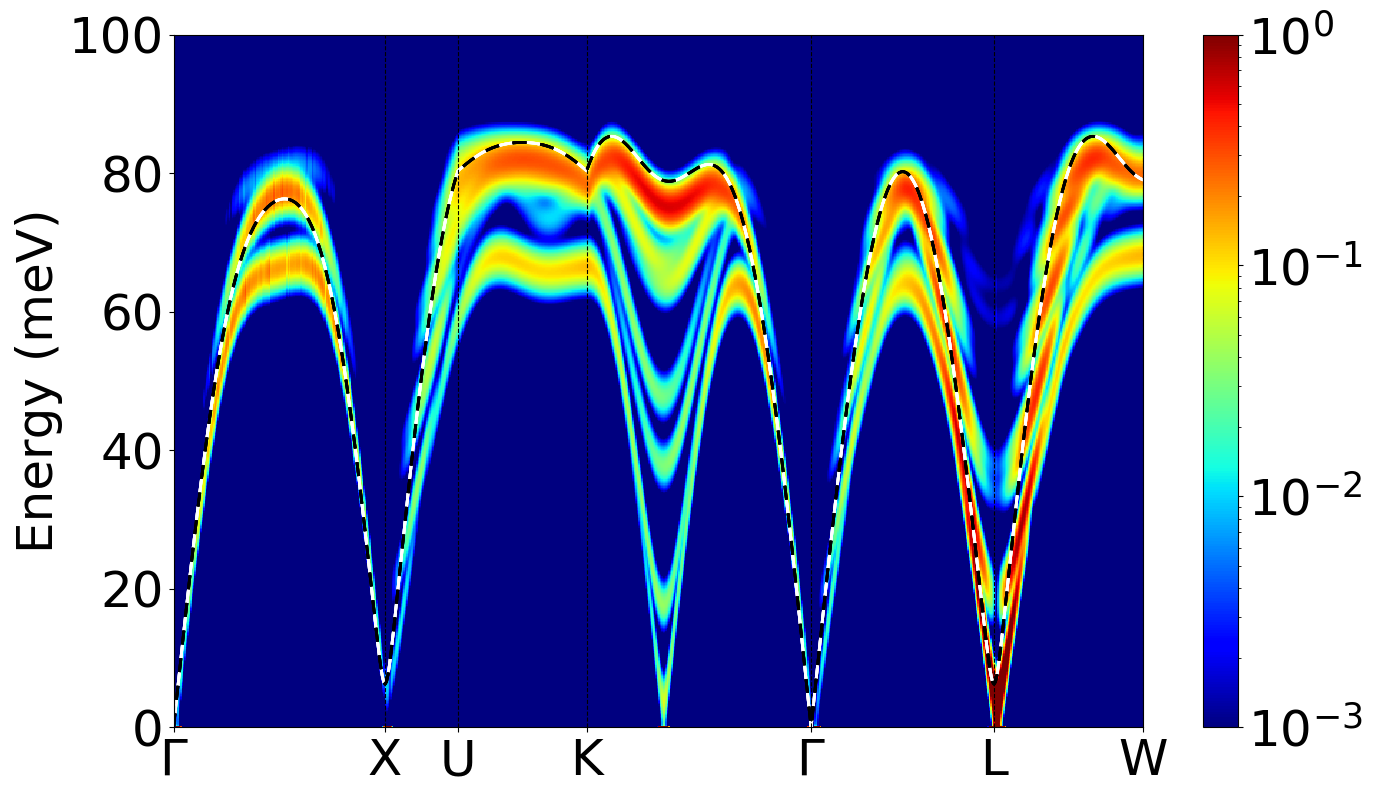}

  \label{fig:NiO_3}
\end{subfigure}

\medskip
\begin{subfigure}{0.315955766\textwidth}
    \raggedright
    (d)
  \includegraphics[width=\linewidth]{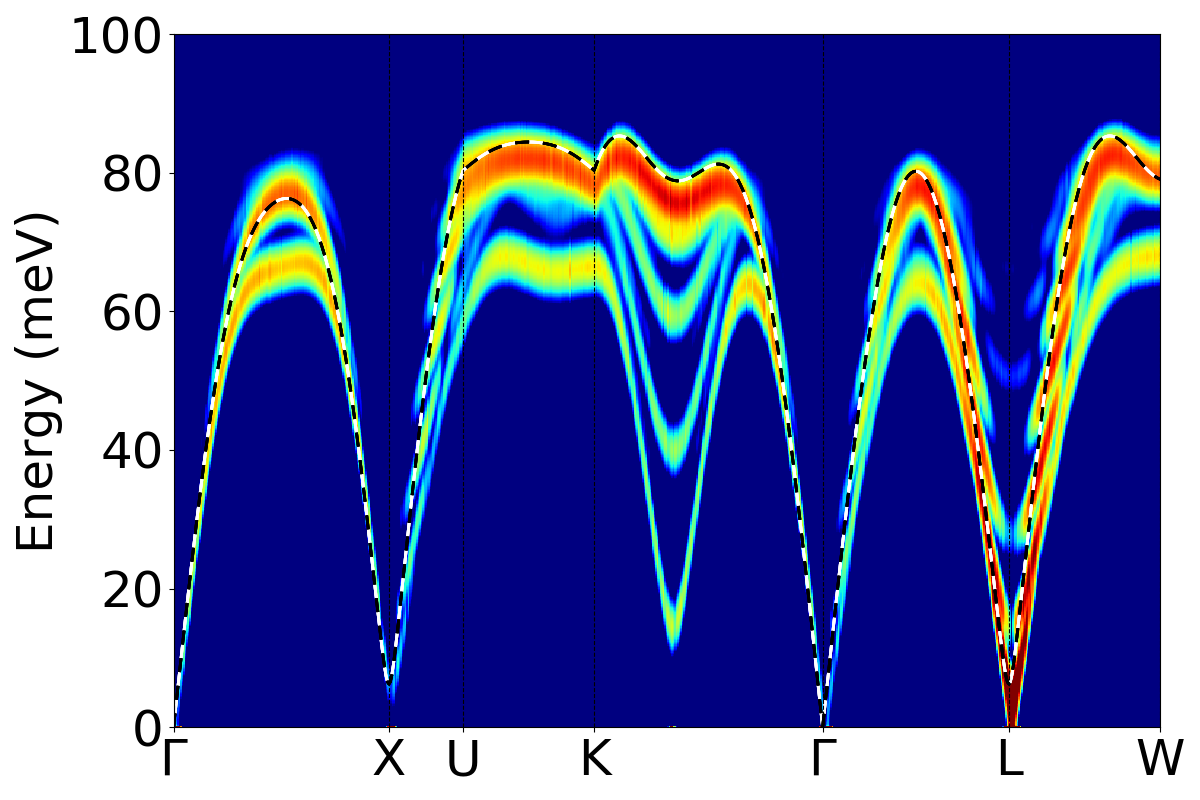}

  \label{fig:NiO_4}
\end{subfigure}\hfil 
\begin{subfigure}{0.315955766\textwidth}
    \raggedright
    (e)
  \includegraphics[width=\linewidth]{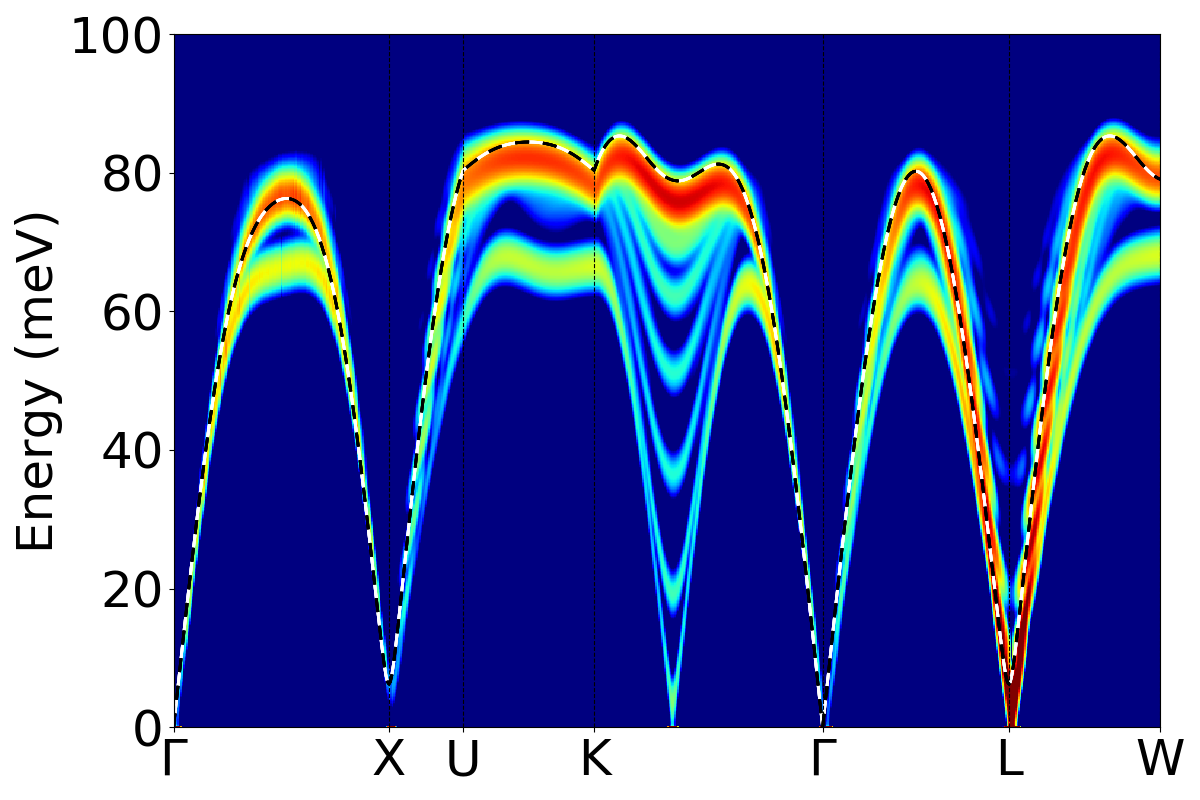}

  \label{fig:NiO_5}
\end{subfigure}\hfil 
\begin{subfigure}{0.368088468\textwidth}
    \raggedright
    (f)
  \includegraphics[width=\linewidth]{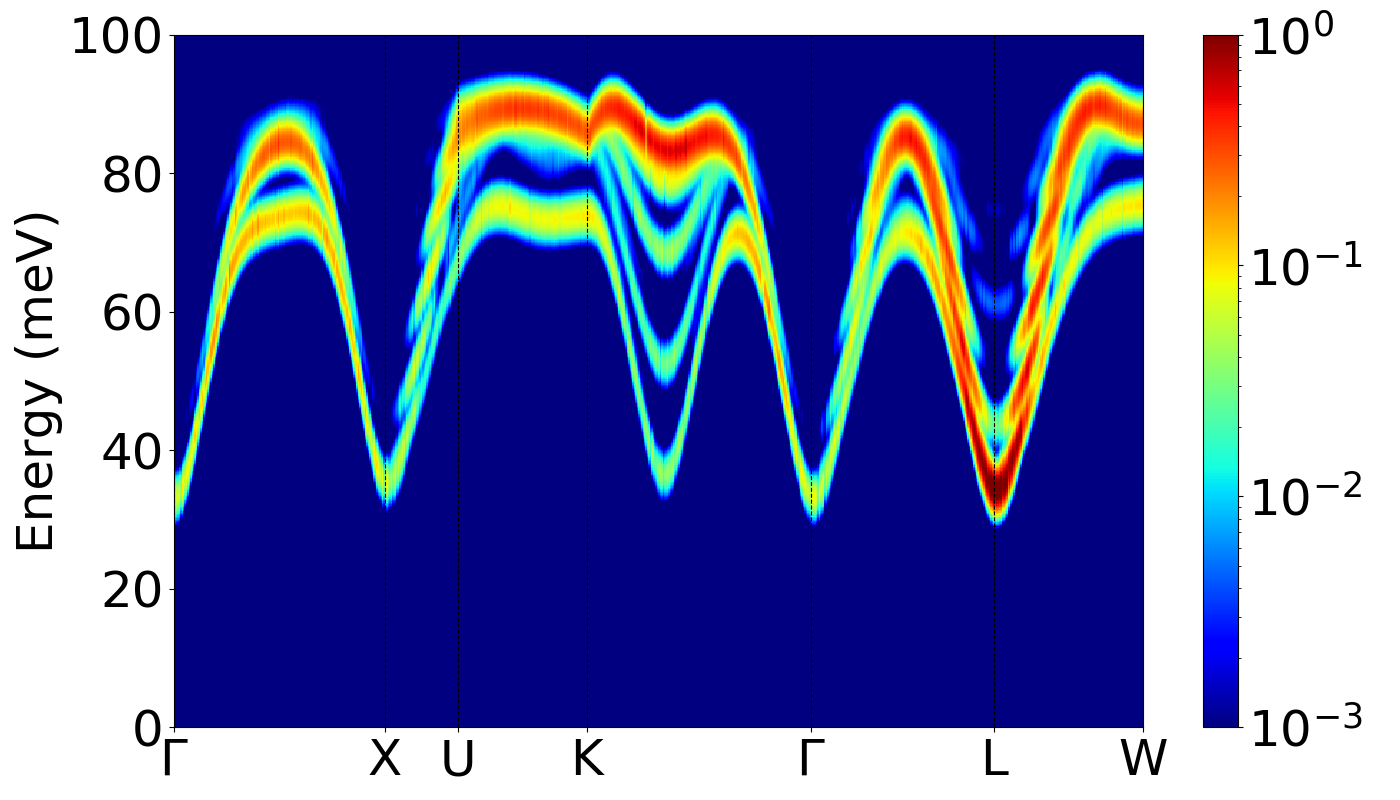}

  \label{fig:NiO_6}
\end{subfigure}
\caption{Spin scattering function of NiO thin films with a) 5, b) 10, c) 15, d) 20, e) 30 monolayers. f) 20 Monolayers with an added 5 meV of anisotropy in the direction of the Néel vector, and a temperature of 300K. }
\label{fig:NiO1}
\end{figure*}

Figure \ref{fig:NiO1} shows the spin scattering function calculated for NiO(100) with 5, 10, 15, 20 and 30 monolayers, alongside the bulk NiO case, again represented by the black/white line. 

As expected and similar to bbc Fe, we confirm the tendency to match the bulk case in our calculations as the number of monolayers is increased. A softer energy mode related to the reduced interaction of the magnetic moments at the surfaces also appears for the thin film cases. 

In Fig \ref{fig:NiO1}(f) we show that adding $K=5$ meV of magnetocrystalline anisotropy in the direction of the magnetic moments in the 20 monolayers case results in a shift in energy for all the modes to higher energies. In contrast to the bcc Fe case, this change is not rigid: the lower energies of the modes are more affected than the higher ones.

\begin{figure}
    \centering
    \includegraphics[scale=0.18]{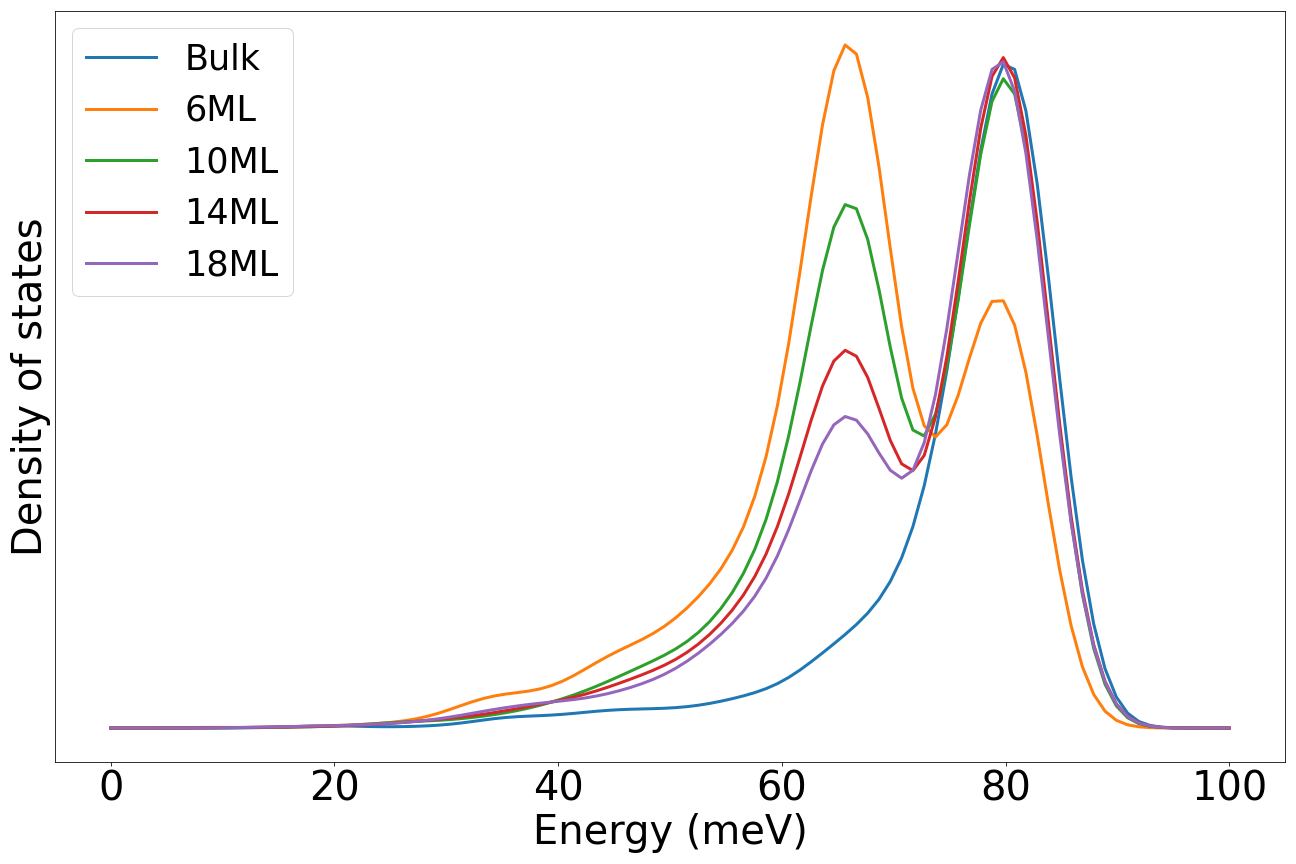}
    \caption{Density of states of NiO comparing Bulk with varying sizes of thin films with a temperature of 300K. }
    \label{fig:NiODOS}
\end{figure}

As with Fe, the magnon DOS of NiO(100) was calculated and is shown in figure \ref{fig:NiODOS}.  The calculation confirms the appearance of the confined modes and their relation to the bulk case. The general tendency is again of a very pronounced confinement-related peak that continuously decreases with the increase of the thin film's size and eventually merges with the bulk-like peaks. We note that in the presented range of DOS calculations, only in ultra-thin films, below 10 ML, are the confined modes comparable to or larger than their bulk-like counterparts. 

\begin{figure*}[h!]
    \centering 
\begin{subfigure}{0.315955766\textwidth}
    \raggedright
    (a)
  \includegraphics[width=\linewidth]{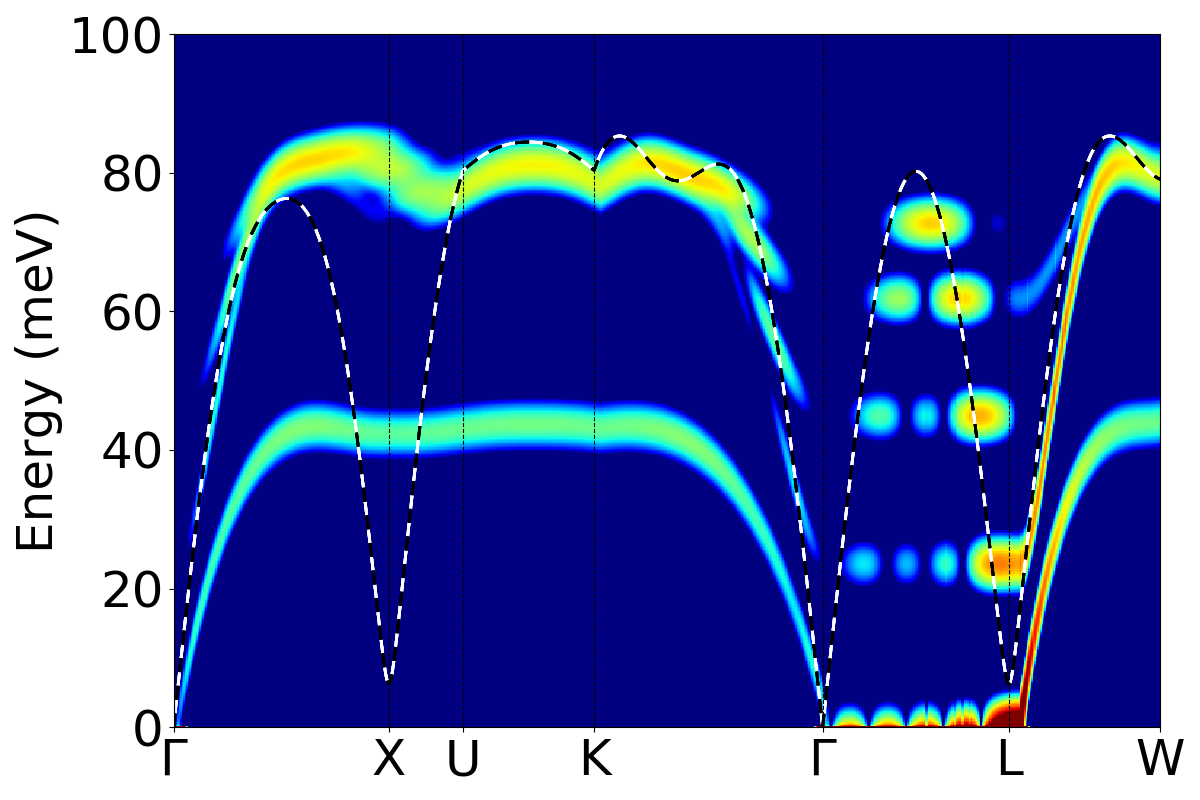}

  \label{fig:NiO1111}
\end{subfigure}\hfil 
\begin{subfigure}{0.315955766\textwidth}
    \raggedright
    (b)
  \includegraphics[width=\linewidth]{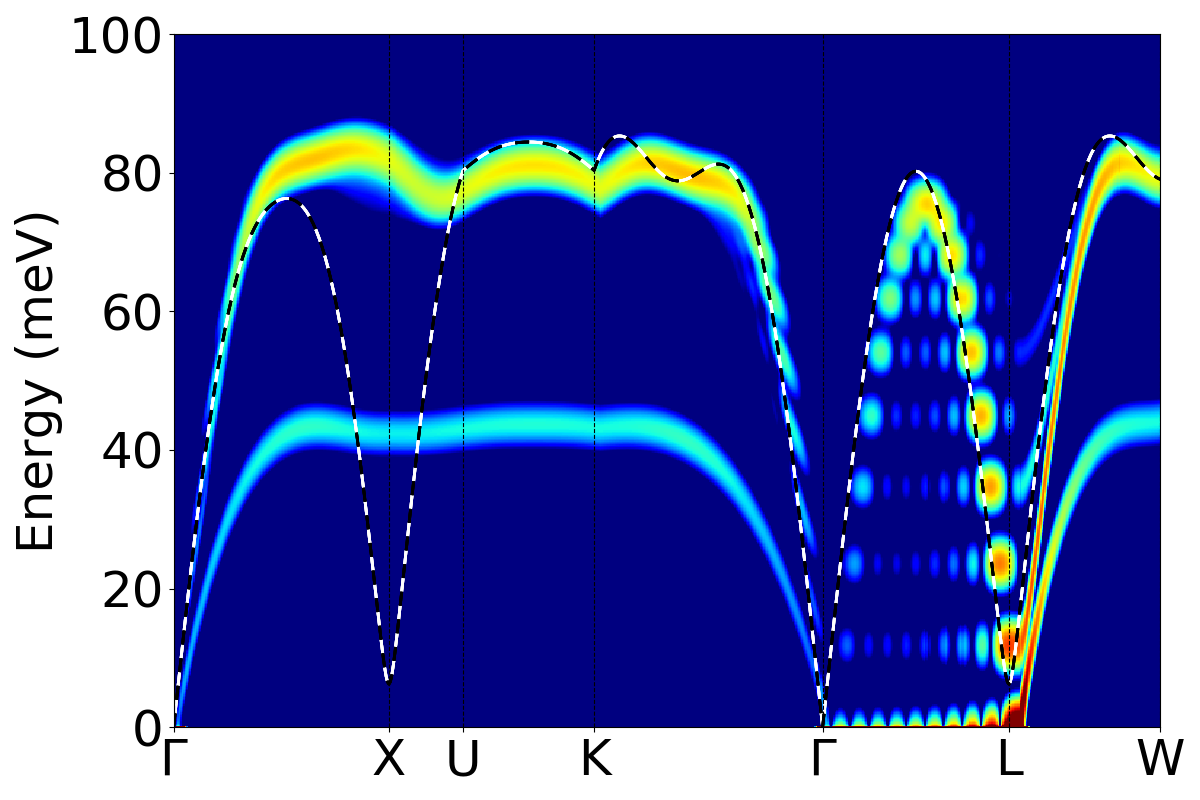}

  \label{fig:NiO1112}
\end{subfigure}\hfil 
\begin{subfigure}{0.368088468\textwidth}
    \raggedright
    (c)
  \includegraphics[width=\linewidth]{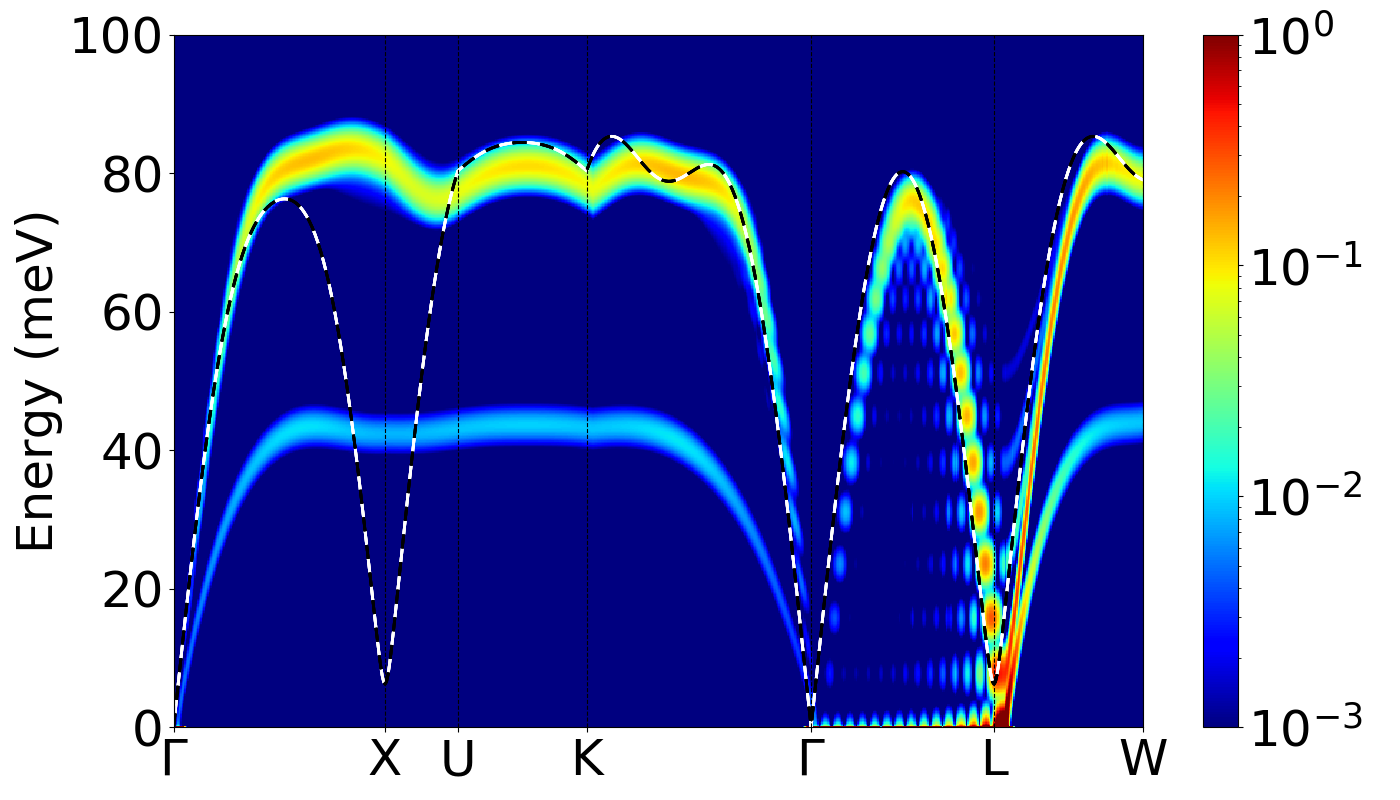}

  \label{fig:NiO1113}
\end{subfigure}

\medskip
\begin{subfigure}{0.315955766\textwidth}
    \raggedright
    (d)
  \includegraphics[width=\linewidth]{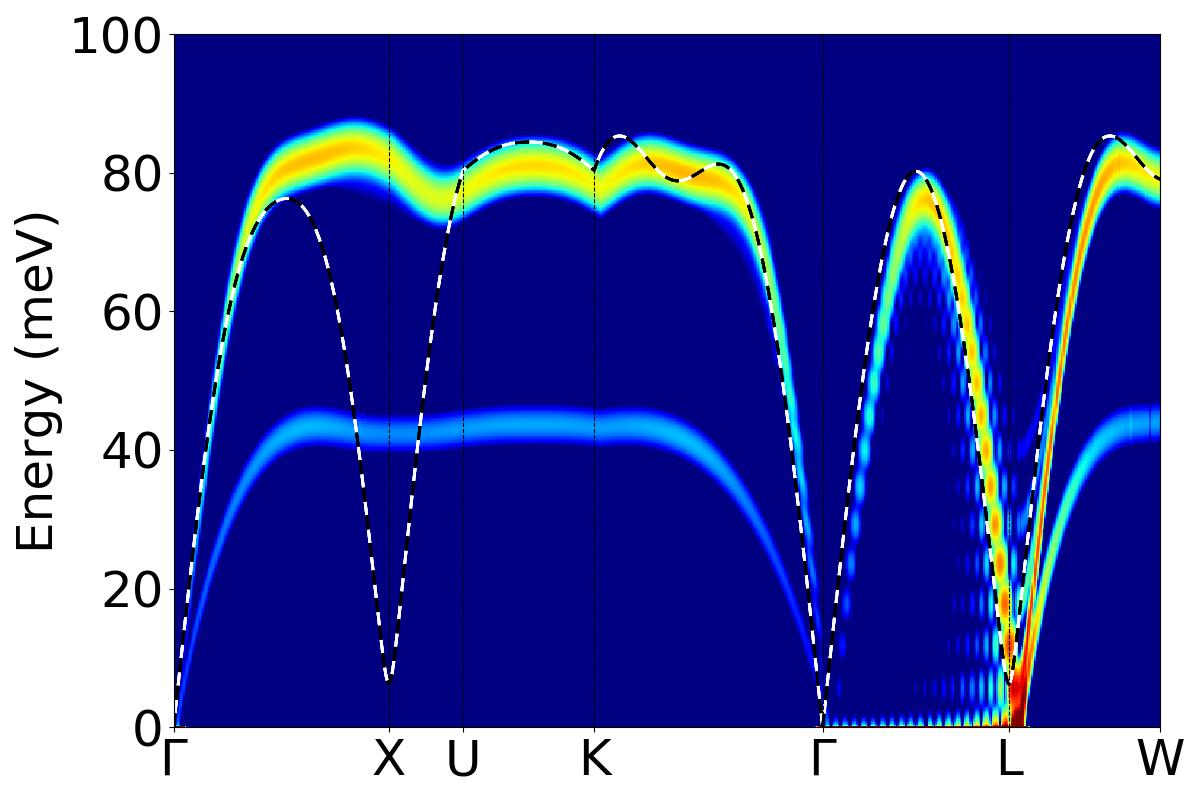}

  \label{fig:NiO1114}
\end{subfigure}\hfil 
\begin{subfigure}{0.315955766\textwidth}
    \raggedright
    (e)
  \includegraphics[width=\linewidth]{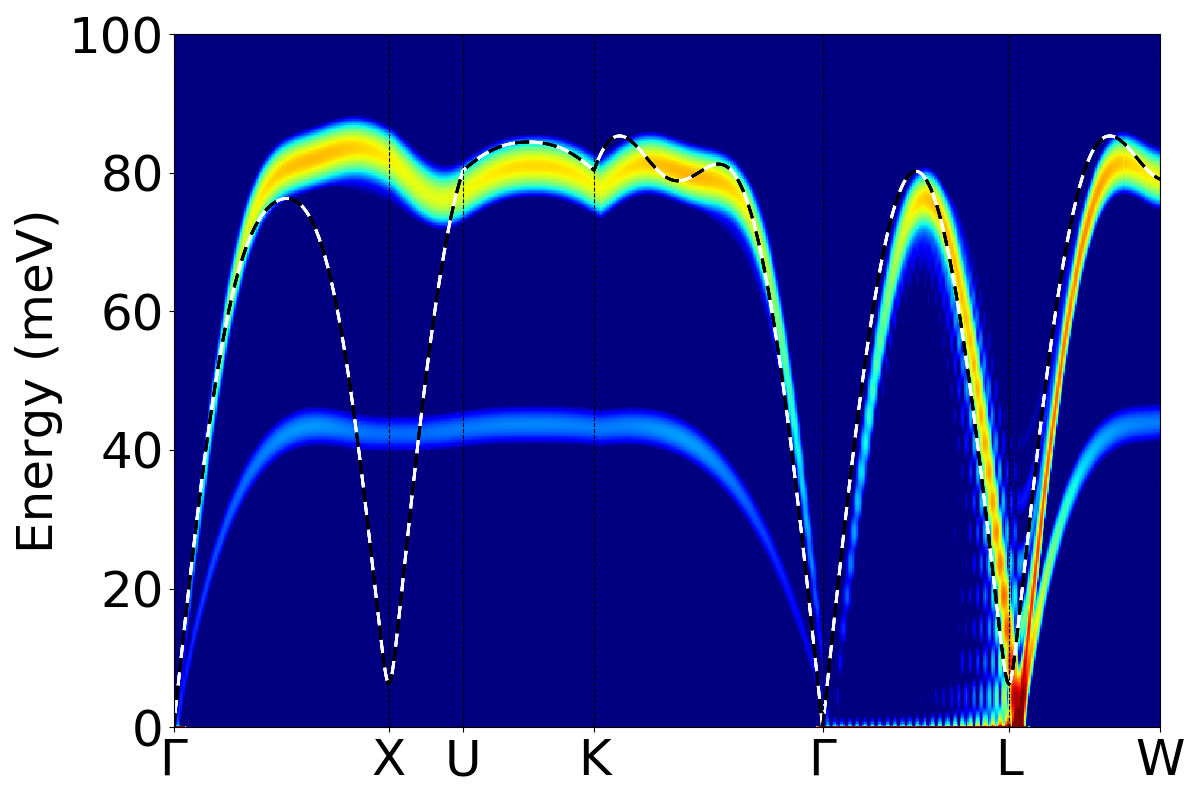}

  \label{fig:NiO1115}
\end{subfigure}\hfil 
\begin{subfigure}{0.368088468\textwidth}
    \raggedright
    (f)
  \includegraphics[width=\linewidth]{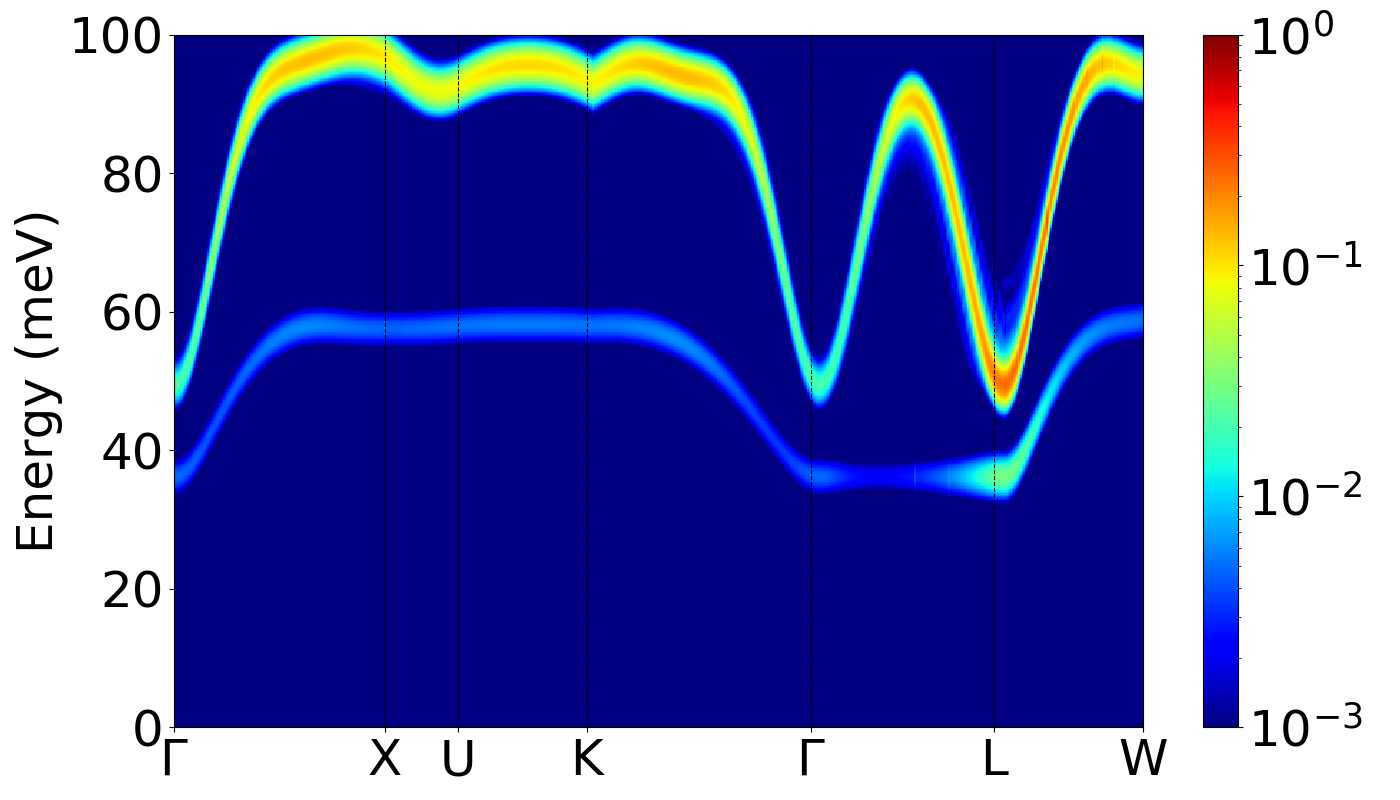}

  \label{fig:NiO1116}
\end{subfigure}
\caption{Spin scattering function of NiO (111) thin films with a)10, b)20, c)30, d)40, e)50 monolayers f) 50 monolayers with an added 10 meV of anisotropy in the direction of the Néel vector, and a temperature of 300K. }
\label{fig:NiO111}
\end{figure*}

The effect of crystallographic direction on confinement is illustrated by a set of complementary calculations carried out on NiO(111) thin films.   Figure \ref{fig:NiO111} shows the spin scattering function of NiO(111) for 10, 20, 30, 40 and 50 monolayers, alongside the bulk spin scattering function.  We also have included the effect of anisotropy in the case of a film of 30 ML with an added anisotropy of 10 meV in the same direction as the Néel vector.

Confinement-induced granularity, due to quasi-momentum quantization, appears in different regions compared to the NiO(100) case, for the chosen path in the Brillouin zone reflecting the crystallographic direction of the confinement. The softer partial interaction mode also appears at lower energies than in NiO(100), with a flatter dispersion. This effect becomes even more apparent when we add an anisotropy in the direction of the Néel vector. A $K$=10 meV anisotropy leads to the hardening of all modes and a separation between the bulk mode and a confined mode at lower energies, and we observe a flat dispersion across the $\Gamma - L$ path corresponding to the out-of-plane direction, which can therefore be interpreted as a surface-confined mode. 

We note also the stark difference between (100) and (111) oriented films in the $\Gamma - X$ direction. While we see a strong dip in the dispersion in the bulk and the (100) films, in the (111) films this dip is not present, showing that the direction of confinement can drastically change some of the features of the inelastic response.

\begin{figure}
    \centering
    \includegraphics[scale=0.18]{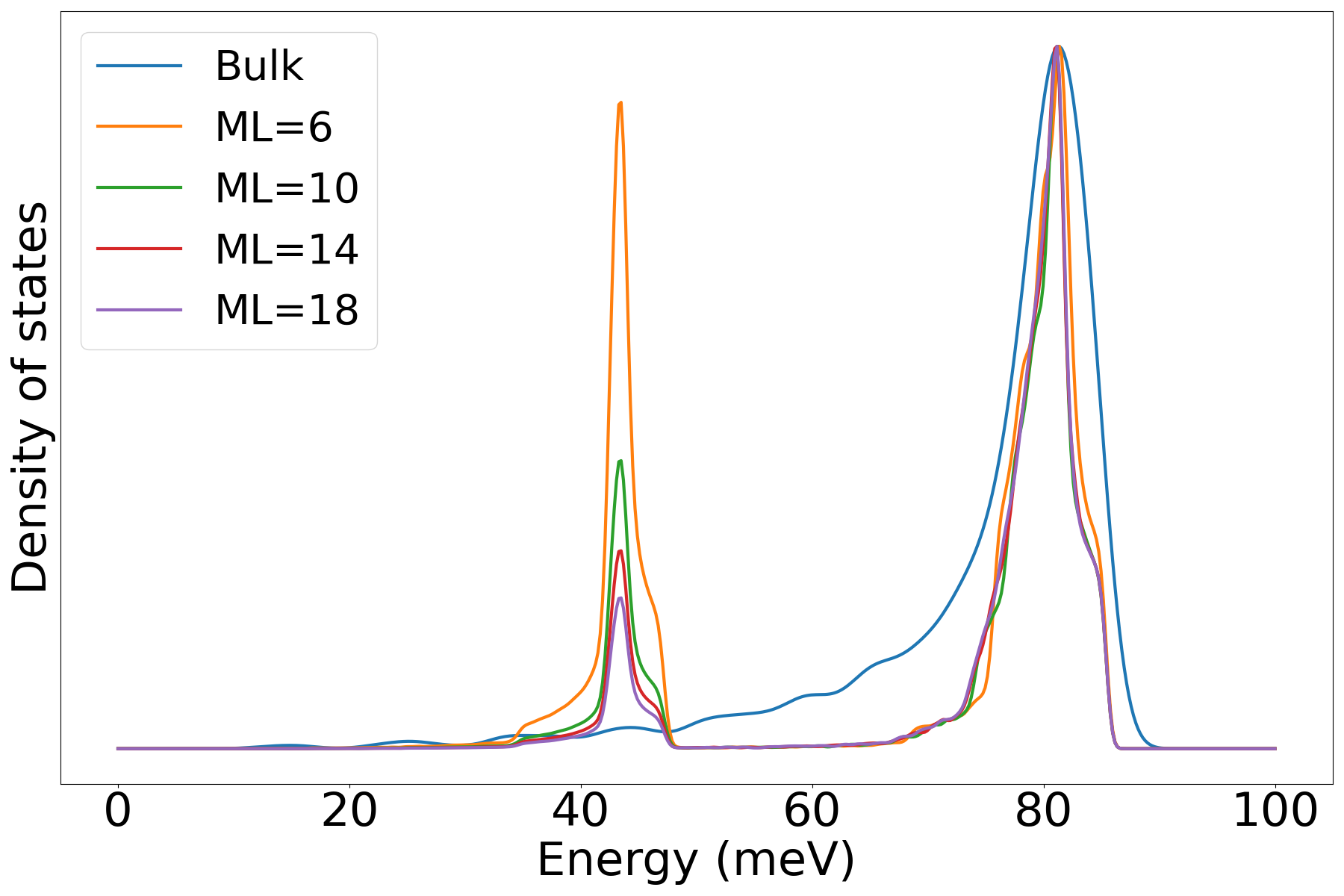}
    \caption{Density of states of NiO (111) comparing Bulk with varying sizes of thin films with a temperature of 300K.  }
    \label{fig:NiODOS111}
\end{figure}

Finally, a comparison of the calculated DOS for NiO(111) for different film thicknesses is given in figure \ref{fig:NiODOS111}. We once more observe the same tendency as in the NiO(100) films, where the magnon DOS peaks due to the surface-related modes are reduced in intensity and merge into the main bulk-like response for (111) films above 10 ML. The main difference with Ni(100) appears in the energy position (42 meV \textit{vs.} 65 meV) of the surface-related peak which is due to the direction of confinement.

\section{Conclusion}\label{Conclusion}

We outline a methodology to calculate the inelastic scattering for magnons relying on the spin scattering function, having used a second quantization approach to obtain a Hamiltonian that allows us to include lower dimensionalities in the systems considered, especially for applications-relevant thin film systems. The method was use to study thin films of bcc Fe(100), and for NiO the (100) and (111) crystallographic orientations. The calculated results have revealed distinct characteristics that emphasize the significance of film thickness and crystal orientation on the magnon modes in these systems. In all cases, we see the emergence of softer modes related to the partial interaction of the magnetic moments close to the surfaces of the material. The appearance of these softer modes related to confinement is evident in the evaluated magnon density of states, where we see the emergence of peaks at lower energies, which we ascribe as being related to confinement. Comparing the magnon density of states for the two different crystallographic orientations for NiO we see that the confinement-related peak appears at different energies, (65 meV for (100) and 42 meV for (111)), showing the importance of the direction in which dimensionality is reduced. The finite crystal size leads to granularity in the spin scattering function across various directions in the Brillouin zone, due to the quantization of the quasi-momenta. This effect is comparable in the two different crystallographic orientations of NiO, where the granularity appears in the path in reciprocal space that is oriented in the direction of the thin film. 
Additionally, we demonstrated the role of adding magneto-crystalline anisotropy to both Fe and NiO films, which led to an overall hardening of the magnon modes. When anisotropy is only included in surface layers, illustrated in the case of bbc Fe, we see a shift of the confinement-related magnon DOS peak to higher energies with the increase of the surface anisotropy.  
Overall, this study contributes to the growing body of knowledge in the field of magnonics and serves as a foundation for future research endeavours aimed at harnessing the unique properties of confined magnon modes for fundamental studies and technological applications.

\section{Acknowledgement}\label{ack}

This project was undertaken on the Viking Cluster, which is a high-performance compute facility provided by the University of York. We are grateful for computational support from the University of York High-Performance Computing service, Viking and the Research Computing team. SuperSTEM is the UK National Research Facility for Advanced Electron Microscopy, supported by the Engineering and Physical Sciences Research Council through grant number EP/W021080/1. We acknowledge financial support from EPSRC via Grant No. EP/V048767/1 and Royal Society Grant No. IES/R1/211016.



\vspace{2cm}

\bibliographystyle{unsrt}

\bibliography{bibliography}

\end{document}